\documentclass[11pt, a4paper]{article}
\usepackage{jheppub}
\usepackage{orcidlink}

\usepackage{epsfig}
\usepackage{amsfonts}
\usepackage{amsmath}
\usepackage{amssymb}
\usepackage{bbm,bm}
\usepackage{caption}
\usepackage{subcaption}
\usepackage{float}
\usepackage{physics}

\usepackage{comment}
\usepackage[normalem]{ulem} 

\def\backtick{\char18}
\usepackage{listings}
\newcommand{\VT}{V^{ }_\rmii{$T$}}

\newcommand\MSbar{$\overline{\rm MS}$}

\newcommand{\rmii}[1]{{\mbox{\tiny\rm{#1}}}}

\newcommand{\Tint}[1]{{\hbox{$\sum$}\!\!\!\!\!\!\!\int\,}_{\!\!\!\!\raise-0.9ex\hbox{$\scriptstyle{#1}$}}}
\newcommand{\Tinti}[1]{{{\Sigma}\!\!\!\!\raise0.3ex\hbox{$\int$}_\rmii{${#1}$}}}
\newcommand{\Tintip}[1]{{{\Sigma'}\!\!\!\!\!\raise0.3ex\hbox{$\int$}_\rmii{${#1}$}}}

\newcommand{\Vone}{V^{(1)}}
\newcommand{\Vtree}{V^{(0)}}
\newcommand{\field}{\varphi}
\newcommand{\fieldh}{h}

\usepackage{pifont}
\newcommand{\daisy}{\selectfont \ding{95}}%

\title{Gravitational waves from supercooled phase transitions:  dimensional transmutation meets dimensional reduction}
			
\author[1]{Maciej~Kierkla \orcidlink{0000-0002-2785-5370},}
\author[1]{Bogumi{\l}a {\'S}wie{\.z}ewska \orcidlink{0000-0003-0169-211X},}
\author[2,3,4,5]{Tuomas V.~I.~Tenkanen \orcidlink{0000-0002-3087-8450}}
\author[6,7]{\\and Jorinde van de Vis \orcidlink{0000-0002-8110-1983}}

\affiliation[1]{Faculty of Physics, University of Warsaw, ul.\ Pasteura 5, 02-093 Warsaw, Poland}
\affiliation[2]{Department of Physics and Helsinki Institute of Physics,
	University of Helsinki, P.O.~Box 64, FI-00014 Helsinki,
	Finland}
\affiliation[3]{Nordita,
	KTH Royal Institute of Technology and Stockholm University,\\
	Hannes Alfv\'ens v\"ag 12,
	SE-106 91 Stockholm,
	Sweden}
\affiliation[4]{Tsung-Dao Lee Institute \& School of Physics and Astronomy, Shanghai Jiao Tong University, Shanghai 200240, China}
\affiliation[5]{Shanghai Key Laboratory for Particle Physics and Cosmology, Key Laboratory for Particle Astrophysics and Cosmology (MOE), Shanghai Jiao Tong University, Shanghai 200240, China}
\affiliation[6]{Institute for Theoretical Physics, Utrecht University,
	Princetonplein 5, 3584 CC Utrecht, The Netherlands}
\affiliation[7]{Instituut-Lorentz for Theoretical Physics, Leiden University, Niels Bohrweg 2, 2333 CA Leiden, the Netherlands}

\emailAdd{maciej.kierkla@fuw.edu.pl}
\emailAdd{bogumila.swiezewska@fuw.edu.pl}
\emailAdd{tuomas.tenkanen@helsinki.fi}
\emailAdd{vandevis@lorentz.leidenuniv.nl}
			
\abstract{
Models with radiative symmetry breaking typically feature strongly supercooled first-order phase 
transitions, which result in an observable stochastic gravitational wave background. In~this work, we 
analyse the role of higher-order thermal corrections for these transitions, applying high-temperature 
dimensional reduction to a theory with dimensional transmutation. In~particular, we study to what 
extent high-temperature effective field theories (3D EFT) can be used. We find that despite 
significant supercooling down from the critical temperature, the high-temperature expansion for the 
bubble nucleation rate can be applied using the 3D EFT framework, and we point out challenges in 
the EFT description. We compare our findings to previous studies and find that the next-to-leading 
order corrections obtained in this work have a significant effect on the predictions for GW 
observables, motivating a further exploration of higher-order thermal effects.
}

\arxivnumber{2312.12413 }

\begin{document}
	\maketitle
	


%
\section{Introduction}
\label{sec:intro}
	
The observation of a stochastic gravitational wave (GW) background from a primordial first-order 
phase transition would unravel information about underlying particle physics beyond that of the 
Standard Model (SM). A very interesting beyond-the Standard-Model (BSM) scenario is the case of 
a supercooled first-order phase transition, which typically arises in models with classical scale 
invariance (or nearly conformal dynamics)~\cite{Randall:2006, Konstandin:2010, 
Konstandin:2011,Hambye:2013,Jaeckel:2016}. In such a case, the phase transition completes at a 
temperature much below the critical temperature. As a result, the potential energy difference 
between the high-temperature and low-temperature phases becomes very large, and the amount of 
energy released -- relative to the radiation energy density -- is orders of magnitude larger than in 
scenarios without significant supercooling. Large energy release results in a strong GW signal 
sourced by the sound waves in the plasma or the collisions of the bubble walls~\cite{Ellis:2019, 
Ellis:2020, Lewicki:2020, Lewicki:2020azd, Lewicki:2022pdb, Kierkla:2022odc}. Predictions of the 
GW spectrum for models with classical scale invariance~\cite{Hambye:2013, Jaeckel:2016, 
Hashino:2016, Jinno:2016, Marzola:2017, Hashino:2018, Baldes:2018, Prokopec:2018,Marzo:2018, 
Mohamadnejad:2019vzg, Kang:2020jeg, Mohamadnejad:2021tke, Dasgupta:2022isg, 
Kierkla:2022odc, Frandsen:2022klh} indicate that the signal could be readily observed by the Laser 
Interferometer Space Antenna (LISA)~\cite{Audley:2017drz} and other next-generation GW 
detectors~\cite{Kawamura:2011zz,Harry:2006fi,Guo:2018npi}. This makes models with classical 
scale invariance and strong supercooling an interesting theoretical playground, and accurate 
predictions of the GW spectrum in terms of the free parameters of such models are essential to 
determine if a potentially observed GW signal was caused by a phase transition in such a model.

Predicting the GW signal requires a determination of thermal parameters describing the phase transition, such as the percolation temperature $T_p$, the strength $\alpha$, the (inverse) time or length scale of the transition, $\beta$ or $R_*$ and the wall velocity $v_w$. In many studies, the phase transition parameters are obtained from the one-loop effective potential at finite temperature, with so-called daisy resummation accounting for a resummation of a class of diagrams enhanced in the infrared (IR) due to 
thermal screening. In recent years, it has become clear in the context of Higgs portal models~\cite{Gould:2021} and the Standard Model Effective Field Theory~\cite{Croon:2020cgk} that this approach might not predict the thermal parameters with sufficient precision, and the corresponding uncertainty of the GW signal can be 
several orders of magnitude. In the work at hand, we apply similar  higher-order thermal corrections to models with classical scale invariance.
	
The reason for the poor convergence of the computation at finite temperature, is that bosonic 
low-energy modes become highly occupied in a thermal plasma. This results in a breakdown of the 
usual loop expansion~\cite{Kapusta:1979fh,Parwani:1991gq,Arnold:1992}. Indeed, the standard 
one-loop procedure suffers from an incomplete treatment of the perturbative expansion, which 
reveals itself as an uncancelled dependence on the renormalisation scale~\cite{Arnold:1992, 
Gould:2021}. The escape out of this distress is the use of a dimensionally reduced effective field 
theory (EFT)~\cite{Ginsparg:1980ef,Appelquist:1981vg,Kajantie:1995dw,Braaten:1995cm}, that is 
constructed to account for thermal scale hierarchies and consistently incorporates the required 
thermal resummations, which significantly reduces the uncertainty of the GW signal 
predictions~\cite{Croon:2020cgk,Gould:2021} (c.f.\ also ref.~\cite{Gould:2023jbz}). This method 
allows one to construct an EFT for only the degrees of freedom that are driving the phase transition 
at IR length scales. The heavy ultraviolet (UV) modes are integrated out, and their effect is captured 
in the parameters of the EFT via matching.  Constructing the EFT can be a technically challenging 
endeavour compared to the use of mere one-loop thermal functions with minimal daisy resummation 
that encode the leading behaviour of the effective potential, but this obstacle has been largely 
removed by {\tt DRalgo}~\cite{Ekstedt:2022bff}, which has automated the matching procedure and 
the computation of the effective potential in the EFT for generic models. Furthermore, the 
formulation in terms of an effective field theory combined with strict perturbative expansions has 
been shown to provide a theoretically sound setup for computations, that is free of residual gauge 
dependence, imaginary parts, spurious IR-divergences or double counting 
contributions~\cite{Gould:2021ccf,Lofgren:2021ogg,Hirvonen:2021zej,Schicho:2022wty,Ekstedt:2022zro,Lofgren:2023sep,Gould:2023ovu}.
 Indeed, in the terminology used in ref.~\cite{Gould:2023ovu} we implement the \textit{mixed 
method} in the computation of the bubble nucleation rate, which is based on the strict expansion for 
the action around the leading order bounce solution. 

So far the dimensionally reduced EFT approach has not been applied to models with classical scale invariance. At first glance, the approach might not even seem suitable for the study of supercooled phase transitions, as the construction of the dimensionally reduced EFT relies on scale hierarchies in a high-temperature (HT) expansion, assuming that the field-dependent masses are small compared to the temperature. This assumption seems not at all appropriate for a phase transition in a scale-invariant model: 
the position of the minimum of the potential of the transitioning field exceeds the temperature by multiple orders of magnitude. This suggests that applying the dimensionally reduced EFT to models with classical scale invariance might do  more harm than good: does the inclusion of higher-order corrections in the effective potential come at the cost of applying the HT expansion in a regime where it is not at all valid? 

In this work, we will argue  that the EFT relying on the HT expansion \emph{can} be used for parts of 
the computation. The crux is that the transitioning field does not transition directly to the minimum 
of the potential, but remains in the regime of validity of the HT approximation. Therefore, along the 
path formulated in refs.~\cite{Gould:2021ccf, Ekstedt:2022zro, Hirvonen:2022jba,Lofgren:2023sep, 
Gould:2023ovu}, we compute the thermal contributions coming from the so-called hard scale (c.f.\ 
section~\ref{sec:HT-EFT}) and construct an EFT for the bubble nucleation at the soft scale. This EFT 
can be used for the  determination of the nucleation and percolation temperature, and the typical 
length scale of the transition. Other parameters, such as the phase transition strength, \emph{do} 
depend on the value of the potential at its minimum. These quantities have to be determined without 
the high-temperature expansion, but follow from the zero- or low-temperature potential (see e.g.\ 
ref.~\cite{Kierkla:2022odc} for the details on how to compute the reheating temperature, or the 
potential energy difference $\Delta V$). For concreteness, we will demonstrate the approach 
explicitly in the SU(2)cSM model~\cite{Hambye:2013, Carone:2013}, a~conformal extension of the 
SM. 
	
We find that the next-to-leading (NLO) corrections included in the EFT modify the predictions for the properties of the phase transition 
significantly, as compared to earlier results based on daisy resummation~\cite{Kierkla:2022odc}. For example, the percolation temperature can change by 100\%, whereas the changes in the length scale, given by the normalised bubble radius $R_*H_*$, reach 50\%. Since the signal is expected to be well visible with LISA, 
it would be possible
to reconstruct the values of $R_*H_*$ and the reheating temperature with good accuracy~\cite{Gonstal:2023}. 
This clearly shows the importance of providing the most precise theoretical predictions possible.
 
Interestingly, the modification of the potential at NLO accounts only for a part of the large correction described above. A correction of the kinetic term in the action, only appearing at NLO, is responsible for a significant shift in the results. This kind of correction is not straightforward to include within the conventional daisy-resummed approach, which shows the importance of using the EFT framework. On the other hand, this correction is also a main source of uncertainty in our computation, and it could indicate the breakdown of the mass hierarchies at the root of the applied EFT. These non-trivial issues in the construction of the EFT for classically conformal theories set the stage for further studies.
	
This article is organised as follows. In section~\ref{sec:supercooling-at-HT} we review the previous knowledge on phase transitions in models with classical conformal symmetry and discuss the applicability of high- and low-temperature approximations. In section~\ref{sec:model} we introduce our concrete BSM model, the SU(2)cSM, for which in section~\ref{sec:HT-EFT} we construct an effective description for bubble nucleation at high temperature, using an EFT at the soft scale. In section~\ref{sec:results} we present our numerical results and we summarise our findings in section~\ref{sec:summary}. For the convenience of the reader, we provide the expressions for the running couplings in appendix~\ref{app:betas} and our implementation of dimensional reduction using {\tt DRalgo} in appendix~\ref{sec:DR-details}.

%
\section{Supercooling at high temperature}
\label{sec:supercooling-at-HT}
	
The equilibrium properties of a high-temperature plasma can be described in Matsubara's imaginary 
time formalism~\cite{Matsubara:1955ws}. In this formalism, fluctuations of several mass scales 
arise: modes with non-zero Matsubara frequency have masses of the order $\pi T$ or higher at 
temperature $T$, while zero modes can have masses which are parametrically smaller $m \sim g 
T$~\cite{Kapusta:2006pm,Laine:2016hma}. Here, the dimensionless coupling $\frac{g^2}{(4\pi)^2} 
\ll 1$ parametrises the hierarchy. Such a hierarchy allows for a HT expansion with respect to 
$\frac{m}{T} \sim g$, and suggests an EFT picture, where an EFT for long-distance IR physics for 
phase transitions is constructed, by integrating out short-distance non-zero modes in the UV. These 
UV modes screen the modes in the IR and generate thermal mass corrections~\cite{Dolan:1973qd}. 
Capturing such effects requires resummation of perturbative expansions. The scalar field zero 
modes  undergo the phase transition, and since they are static and live in three spatial dimensions, 
this procedure is called high-temperature dimensional reduction. We describe this in more detail in 
section~\ref{sec:HT-EFT}. 

In models with classical scale invariance, all fields are massless at classical level, and massive 
modes are generated radiatively at loop level by quantum corrections. This is called dimensional 
transmutation~\cite{Coleman:1973}. Physical masses depend on the vacuum expectation value of 
the scalar field which is typically much larger than the nucleation temperature, thus $\frac{m}{T} \gg 
\pi$, seemingly invalidating the use of HT expansion.
	
Hence, at first sight, supercooling seems incompatible with the formalism of dimensional reduction 
as the latter relies on the HT expansion. In this section, we delve into this seeming contradiction to 
formulate a consistent prescription for treating supercooled phase transitions with due accuracy. For 
the current purpose, we will phrase our discussion in terms of the one-loop effective potential with 
Arnold-Espinosa -- or daisy -- resummation~\cite{Arnold:1992}: a framework which is familiar to 
most readers and corresponds exactly to dimensional reduction at leading order (see 
section~\ref{sec:compare3D4D}), where thermal corrections to the masses are computed at 
one-loop, and resummed to all orders. We start with a brief review of the temperature-dependent 
effective potential and the parameters characterising phase transitions.
	
%
\subsection{Perturbative description of a phase transition}
\label{sec:PT}
	
In perturbation theory, the effective potential is the central object for computing the properties of the phase transition. It is given by a loop expansion as
\begin{equation}
\label{eq:Veff-1loop}
V=\Vtree + \Vone +\VT,
\end{equation}
where $\Vtree$ corresponds to the tree-level potential, $\Vone$ is the one-loop Coleman-Weinberg correction, including counterterms to remove divergences, and $\VT$ contains thermal corrections, at one loop. Herein, two-loop corrections are not considered. In this work, we will use Landau gauge and the \MSbar\ renormalisation scheme.

The one-loop zero-temperature correction is given by the well-known formula~\cite{Coleman:1973} being a sum of contributions from different fields
\begin{equation}
\Vone(\field)=\frac{1}{64 \pi^2}\sum_{a}n_a M_a^4(\field)\left(\log\frac{M_a^2(\field)}{\mu^2}-C_a\right),
\end{equation}
with $n_a$ counting the number of degrees of freedom as
\begin{equation}
n_a=(-1)^{2s_a} Q_a N_a (2s_a+1),
\end{equation}
where $s_a$ denotes the spin of a given particle, $Q_a=1(2)$ for neutral (charged) particles, and $N_a=1(3)$ for uncoloured (coloured) particles. $C_a=3/2$ for scalars and fermions, and $C_s=5/6$ for vector bosons. Here, for simplicity, we assume the effective potential to be a~function of a single field but it can be straightforwardly generalised to the multi-field case.
	
The one-loop thermal correction is given by
\begin{equation}
\VT(\field, T)=\frac{T^4}{2\pi^2}\sum_{a} n_a J_{T,b/f}\left(\frac{M_a(\field)}{T}\right),
\end{equation}
where the thermal function is defined as 
\begin{equation}
J_{T,b/f}^{}(y) =  \int_0^\infty {\dd} x\, x^2 \log(1\pm e^{-\sqrt{x^2 + y^2}}),
\end{equation}
with the ``$+$'' sign for fermions and ``$-$'' sign for bosons~\cite{Laine:2016hma}. When the temperature is high with respect to the mass scale, $M_{a}/T\ll 1$, 
the thermal functions can be expanded as 
\begin{align}
\label{eq:T-master-high-T}
J^{y \ll 1 }_{T,b}(y)= &\ -\frac{\pi^4}{45} + \frac{\pi^2}{12} y^2 - \frac{\pi}{6} y^{3} - 
\frac{1}{32} y^4\left( \log\frac{y^2}{16 \pi^2} -\frac{3}{2} + 2 \gamma_E \right) + \nonumber\\
&+ \pi^2 y^2 \sum_{i=2}^{\infty}\left(-\frac{1}{4 \pi^2} y^2 \right)^
i \frac{(2 i - 3)!! \zeta(2 i - 1)}{(2 i)!! (i + 1)}, \nonumber \\
J^{y\ll 1}_{T,f}(y)=&\   \frac{7\pi^4}{360} - \frac{\pi^2}{24} y^2 - \frac{1}{32} y^4 \log\left(\frac{y^2}{\pi^2}-\frac{3}{2} + 2 \gamma_E\right) + \nonumber\\
&+\pi^2 y^2 \sum_{i=2}^{\infty}\left(\frac{-1}{4 \pi^2} y^2\right)^i \frac{(2 i - 3)!! \zeta(2 i - 1)}{(2 i)!! (i + 1)} \left(2^{2 i - 1} - 1\right),
\end{align}
and in the opposite regime $M/T\gg 1$, the expansion is
\begin{align}
J_{T}^{y\gg 1}(y)= &-e^{-y} \left(\frac{\pi}{2} y^{3}\right)^{\frac{1}{2}}
\sum_{i=0}^{\infty}\frac{1}{2^i i!} \frac{\Gamma(\frac{5}{2} + i)}{\Gamma(\frac{5}{2} - i)} y^{-i},\label{eq:low-T}
\end{align}
which is the same for bosons and fermions. Above, $\gamma_E$ denotes the Euler-Mascheroni 
constant, $\zeta$ is the Riemann zeta function and $\Gamma$ the gamma function.
	
At high temperature, diagrams that involve IR-sensitive zero modes get enhanced due to thermal 
screening, and these diagrams need to be resummed. For now, we will employ the Arnold-Espinosa 
daisy resummation method~\cite{Arnold:1992}, and further sections will be devoted to a detailed 
discussion of applying the dimensional reduction scheme. The one-loop potential with daisy 
resummation reads
\begin{equation}
\label{eq:Veff-1loop-daisy}
V=\Vtree + \Vone + \VT +V_{\textrm{daisy}},
\end{equation}
where
\begin{equation}
\label{eq:daisy}
V_{\textrm{daisy}}(\field,T)=-\frac{T}{12\pi}\sum_i n_i \left[(M_{i,\textrm{th}}^2(\field,T))^{3/2}-(M_i^2(\field))^{3/2}\right],
\end{equation}
and $M_{i,\textrm{th}}^2(\field,T)$ denotes the thermally corrected mass squared, which is the sum of the squares of the zero-temperature and the thermal mass. Diagrammatically, the first term here is a sum of an infinite number of diagrams of a one-loop zero-mode diagram with non-zero-mode one-loop diagrams attached around it \daisy, hence the name daisy. The physics behind this construction is clear: the non-zero UV modes screen the zero mode living at the IR scale, and the first term in eq.~\eqref{eq:daisy} is nothing but a result of a one-loop diagram of the zero-mode, with resummed mass. The second term merely removes the double counting, since the zero mode contribution with unresummed mass is already included in the cubic term of eq.~\eqref{eq:T-master-high-T}. We emphasise that eq.~\eqref{eq:daisy} should only be added to the effective potential whenever the HT expansion is valid, as it is an essential assumption in its derivation. Using the 3D EFT approach, it is straightforward to derive eq.~\eqref{eq:daisy}, and we compute it explicitly in section~\ref{sec:HT-EFT}.  
	
The temperature evolution of the effective potential determines the details of the phase transition. A 
supercooled phase transition typically proceeds as follows. At high temperature, the scalar field 
fluctuates around the symmetric minimum, and as the temperature decreases, another minimum is 
formed. At the critical temperature $T_c$ the two minima become degenerate and at lower 
temperatures the symmetry-breaking minimum becomes energetically favourable. It is characteristic 
of supercooling that the transition does not proceed right after it has become energetically 
favourable. First, at temperature $T_V$, the Universe enters a stage of thermal inflation induced by 
the large amount of energy stored in the false vacuum. Then, at some point, the field transitions to 
the true ground state, due to getting kicked by thermal fluctuations (typically quantum tunnelling is 
much less probable~\cite{Baldes:2018, Kierkla:2022odc}). The nucleation temperature, $T_n$, at 
which at least one bubble of the true vacuum is nucleated per Hubble volume is considered as the 
onset of the transition. Later the bubbles percolate at the percolation temperature, $T_p$.  To 
consider the phase transition complete, not only the fraction of the volume turned into the true 
vacuum has to be big enough, but also the volume of the false vacuum should be shrinking at 
$T_p$~\cite{Ellis:2018}, see also ref.~\cite{Athron:2022mmm}. This condition is not trivially satisfied 
for transitions taking place during a phase of thermal inflation and 
thus it constrains the available parameter space. The size of the bubbles at the moment of collision, $R_*$, can be used to estimate the length scale of the transition. 
It can be used interchangeably with the (inverse) time scale of the transition, $\beta_*$, given by the derivative of the decay rate of the false vacuum. During the phase transition latent heat is released, which in the case of supercooling is tightly related to the strength of the transition $\alpha$ given by 
\begin{align}
\alpha \simeq \frac{\Delta V}{\rho_{\rm rad}},
\end{align}
where $\rho_{\rm rad}$ is the energy density stored in radiation and $\Delta V$ is the potential 
energy difference between the false and true minima. The released energy is partially converted to 
gravitational waves. In the case of strong supercooling, two production mechanisms -- via bubble 
collisions and sound waves in the plasma -- can be effective (see e.g.\ 
refs.~\cite{Ellis:2020,Kierkla:2022odc}). For the calculation of the terminal Lorentz factor of the 
bubble wall one needs to consider the pressure difference across the wall, and for the NLO pressure 
contribution we use $\gamma$-scaling  
 \cite{Bodeker:2009, Bodeker:2017, Gouttenoire:2021kjv} (for further details, see \cite{Kierkla:2022odc}). We will use the efficiency factor for production via sound waves, $\kappa_{\mathrm{sw}}$ to determine the dominant source, as the efficiency for production via bubble collisions is given by $\kappa_{\mathrm{col}}=1-\kappa_{\mathrm{sw}}$. 
However, most of the energy goes back to the plasma, reheating it back to $T_V$.%
\footnote{
This happens unless the transitioning field is very weakly coupled to the SM plasma. In ref.~\cite{Kierkla:2022odc} it was shown that reheating is efficient in the whole allowed parameter space of the model considered in this work.
} 
The reheating temperature and the length/time scale of the transition (evaluated at $T_p$) are the parameters most relevant for the determination of the resulting GW spectra. Since the GW spectrum depends on $\alpha/(\alpha +1)$ and $\alpha \gg 1$ for strongly supercooled phase transitions, the exact value of $\alpha$ becomes irrelevant. More detailed definitions of the relevant parameters listed above can be found in ref.~\cite{Kierkla:2022odc}, the approach of which we follow here.	
	
The decay  of the false vacuum is controlled by the rate given by%
\footnote{
In some parts of the literature, the action $S(T)$ appears as $S_3/T$, such that $S_3$ has dimension of $T$.
}
\begin{equation}
\Gamma(T)\approx A e^{-S(T)}, 
\end{equation}
where $A$ is $T$-dependent pre-factor.
The three-dimensional Euclidean action is evaluated at the so-called bounce configuration
\begin{equation}\label{eq:Sbounce}
S(T)=\frac{4\pi}{T}\int_0^\infty \mathrm{d}r ~r^2 \left[ \frac{1}{2}\left(\frac{\mathrm{d}\field_b}{\mathrm{d}r}\right)^2 + V(\field_b,T) \right],
\end{equation}
which corresponds to the solution of the bounce equation 
\begin{equation}\label{eq:BounceEOM}
\frac{\mathrm{d}^2\field_b}{\mathrm{d}r^2}+\frac{2}{r}\frac{\mathrm{d}\field_b}{\mathrm{d}r}=\frac{\mathrm{d}V(\field_b,T)}{\mathrm{d}\field_b},
\end{equation}
with boundary conditions $\frac{\mathrm{d}\field_b}{\mathrm{d}r}=0$ for $r=0$ and $\field_b\to0$ for $r\to\infty$.
	
\subsection{High- and low-temperature regimes}
\label{sec:HT-LT-regimes}
	
The ratios of the field-dependent masses to the temperature determine whether the HT or LT limit should be considered. Large field values correspond to large masses and thus LT, while small field values correspond to the HT limit. In models with classical scale invariance, which feature supercooled phase transitions, the scales associated with the global minimum of the potential (the location of which determines the strength of the transition) and with the location of the barrier (where the tunnelling takes place) are widely spread. 
Therefore, we cannot use just one of the limits, either  LT or HT, to have the full picture of the transition.
	
In classically scale-invariant models, the potential around the global minimum, for temperatures below the critical temperature, is in the low-temperature regime. This means that we can use the one-loop thermally corrected potential, without daisy resummation to compute the temperature at which thermal inflation starts, $T_V$, and to compute the vacuum energy close to the nucleation temperature we can even neglect the thermal corrections. 

In the presence of various energy scales, we should use the renormalisation group (RG) improved effective potential to resum the field-dependent logarithmic terms and make the potential perturbative over a wide range of field values. For theories in which the one-loop corrections to the potential are dominated by a single mass scale $M(\field)$ this is straightforward to attain by the field-dependent choice of RG scale, $\mu=M(\field)$. Accordingly, all couplings are run to this scale. However, going from the high-field regime to lower field values, which are relevant for the tunnelling, at some point the ratio $M(\field)/T$ becomes small, which signals the onset of the high-temperature regime.
	
In the high-temperature regime we can use the expansion of eq.~\eqref{eq:T-master-high-T} in the potential but also resummations of higher-order terms are obligatory. The one-loop potential in the high-temperature limit reads (for simplicity we consider here models with bosons only)
\begin{align}
\label{eq:HT-potential}
V&=\Vtree + \Vone 
+\VT+V_{\textrm{daisy}}=\nonumber\\
&=\Vtree + \frac{1}{64 \pi^2}\sum_{a}n_a M_a^4(\field)\left[\log \Big( \frac{(4\pi e^{\gamma_E} T)^2}{\mu^2} \Big)-C_a+\frac{3}{2} \right]\nonumber\\ 
&\phantom{=}+\frac{T^4}{6\pi}\sum_a\left(-\frac{\pi^3}{15}+\frac{\pi}{4}\frac{M_a^2}{T^2}-\frac{1}{2}\frac{M_a^3}{T^3}\right)\nonumber\\
&\phantom{=}-\frac{T}{12\pi}\sum_a n_a \left[(M_{a,\textrm{th}}^2(\field,T))^{3/2}-(M_a^2(\field))^{3/2}\right].
\end{align}
Note that the dependence on the logarithm of the field-dependent mass cancelled 
out.%
\footnote{
It is straightforward to see that a similar cancellation takes place for fermionic fields.
} 
Now the only logarithm present is of the ratio of the temperature and the renormalisation scale. It is thus clear, that in order to preserve perturbativity of the computations one should fix the renormalisation scale to be proportional to the temperature, with some $\mathcal{O}(1)$ proportionality factor. The most natural choice is $\mu=4\pi e^{-\gamma_E} T$, which cancels the logarithmic term entirely. Nonetheless, any choice of $\mu\sim \kappa T$, where roughly $\kappa\in(1,\ 2\pi)$ is acceptable.

This choice builds a bridge between the HT and LT regimes. In the HT regime we have $\mu=\kappa T$, whereas in the LT $\mu=M(\field)$. In the intermediate regime, we thus have $\mu\approx\kappa T\approx M(\field)$: deep in the HT regime we should have $M(\field)/T\ll 1$ and we expect that the breakdown of the applicability of the HT approximation occurs for $M(\field)/T\sim \mathcal{O}(1)$. These observations teach us how to treat the RG-improved potential for the sake of phase transition-related computations: at large field-values the scale should follow the field, whereas at low field-values the scale should be set by the temperature as\footnote{A similar approach was used in ref.~\cite{Kierkla:2022odc}, however, there the thermal cutoff on the running was only introduced in the thermal part of the potential.}
\begin{equation}
\label{eq:mu-choice-1}
\mu=\max(M(\field), \kappa T).
\end{equation}
The thermal cutoff on the running of the couplings prevents them from reaching the Landau poles of e.g.\ the top Yukawa coupling at small field values so it regulates the behaviour of the potential around $\field=0$.
	
One should note, however, that the HT effective potential of eq.~\eqref{eq:HT-potential} is not renormalisation-scale independent, in contrast to the zero-temperature effective potential~\cite{Arnold:1992,Gould:2021}. While the running of the parameters in $\Vtree$ cancels the explicit logarithms appearing in $\Vone$, there is still RG-scale dependence leftover. The implicit running of the term 
\begin{align}
\frac{T^4}{6\pi}\sum_a\left(\frac{\pi}{4}\frac{M_a^2}{T^2} \right),
\end{align}
that corresponds to the thermal correction to the mass at one-loop, is of the same order as the running of $\Vtree$, yet there are no $T^2$-dependent logarithms of the form of
\begin{align}
\frac{T^2}{(4\pi)^2} \log \Big( \frac{(4\pi e^{\gamma_E} T)^2}{\mu^2} \Big),
\end{align}
in the potential for compensation. Indeed, at HT such terms only appear at two-loop order, due to the relatively slower convergence of perturbation theory, induced by enhancement due to thermal screening. The running of the other thermal contributions, in particular from the daisy resummation, is of higher order. In perturbation theory, renormalisation-scale dependence can be used to estimate the size of the missing corrections,%
\footnote{
Or, putting it differently, convergence of perturbation theory is demonstrated when higher orders indeed have a reduced sensitivity to a variation of the renormalisation scale. This is particularly clearly illustrated in figure~2 of ref.~\cite{Ghisoiu:2015uza} that computes the Debye mass of hot Yang-Mills theory at three-loop order. 
} 
and omission of such corrections is the source of one of the largest uncertainties in predictions of GW signals originating from phase transitions~\cite{Croon:2020cgk}. Dimensional reduction will be the technique advocated in this work to be used in the HT regime to include these missing large corrections. In section~\ref{sec:HT-EFT} we discuss how these missing corrections are included and further resummations are performed with care. 
	
In summary, due to different properties of the theory in the UV (large field values, low temperature) and in the IR (small field values, high temperature), the field space naturally divides into two parts:
\begin{enumerate}
\item The low-temperature (LT) regime, where no resummations in the thermal part of the potential are needed. The RG-improved potential with running couplings and fields should be used and the renormalisation scale should follow the value of the field.
\item The high-temperature (HT) regime, where the scale at which computations are performed is set by the temperature. This is the region where thermal resummations are inevitable and the dimensionally reduced theory can be used to cancel renormalisation-scale dependence and systematically include higher-order corrections.
\end{enumerate}
	
In the case of scale-invariant potentials, which we treat as models for supercooled transitions, there is also a natural division of parameters which need to be computed in different field- or scale-regimes:
\begin{enumerate}
\item LT regime: the location of the symmetry-breaking minimum, the energy of the true vacuum (needed to determine the phase transition strength $\alpha$), the temperature at which thermal inflation starts $T_V$, the reheating temperature, $T_r$.
\item HT regime: the percolation and nucleation temperatures, the size of bubbles at collision $R_*H_*$, the inverse time scale of the transition $\beta/H_*$ (normalised to the Hubble length/time).
\end{enumerate}
Therefore, one can compute the quantities in the first category using a finite-temperature RG-improved potential without resummations, while for the quantities in the second category, which are related to solving the bounce equation, the use of the dimensionally reduced theory is required. Below, when we consider a concrete model, we will demonstrate that indeed the escape point for the bounce trajectory lies within the HT region. In the literature the HT approximation has been used for computations related to the phase transition in supercooled transitions, see e.g.\ refs.~\cite{Levi:2022bzt, Salvio:2023qgb, Salvio:2023ynn}. However, using dimensional reduction in the HT regime and relating it to the low-temperature limit through the RG running and using both of them is a novel approach which we present in this work.
		
We will demonstrate how this construction works by applying it to a model with classical scale invariance and an extra SU(2)$_X$ gauge group in the proceeding sections.

\section{Introducing the model}
\label{sec:model}

In this section, we will apply the discussion of the previous section to a concrete BSM model, the so-called SU(2)cSM model~\cite{Hambye:2013, Carone:2013}. It is an extension of the conformal version of the SM (without the explicit mass term for the Higgs field) with a new, ``dark'' SU(2)$_X$ gauge group and a scalar that is a doublet under this new symmetry, while transforming as a singlet under the SM gauge group. This model has been studied extensively in the literature~\cite{Hambye:2013, Carone:2013, Khoze:2014, Pelaggi:2014wba, Karam:2015, Plascencia:2016, Chataignier:2018, Hambye:2018, Baldes:2018, Prokopec:2018, Marfatia:2020, Kierkla:2022odc}, in particular in ref.~\cite{Kierkla:2022odc} a thorough analysis of the thermal history of the Universe within this model has been performed using RG improvement and daisy resummations. In this work, our aim is to improve these results by including higher-order thermal corrections obtained with the dimensionally reduced theory.
	
\subsection{Model at zero temperature}
\label{sec:model-at-zero-T}
	
The model contains two complex scalar doublet fields. We exploit the symmetries to rotate the fields such that the vacuum expectation values are only non-zero in one direction of the field space for each doublet. Then we can write the tree-level potential for the scalar background fields as 
\begin{equation}
\Vtree(\fieldh,\field)=\frac{1}{4}\left(\lambda_{\fieldh} \fieldh^4 + \lambda_{\fieldh \field} \fieldh^2 \field^2 + \lambda_{\field} \field^4\right).\label{eq:Vtree}
\end{equation}
In principle, there are two independent field directions, however, as was discussed in the literature~\cite{Baldes:2018, Prokopec:2018, Kierkla:2022odc}, the tunnelling proceeds along the direction of the new scalar field $\field$ and subsequently the Higgs field $h$ rolls to the true vacuum. Therefore, in our analysis we will focus solely on the $\field$-direction. 

In the one-loop correction to the effective potential, the dominant contribution comes from the dark gauge bosons, $X_\mu$. Therefore, the one-loop potential along the $\field$ direction at zero temperature reads 
\begin{equation}
\Vtree(\field)+\Vone(\field)=\frac{1}{4}\lambda_{\field} \field^4+\frac{9M^4_X(\field)}{64\pi^2}\left(\log\frac{M^2_X(\field)}{\mu^2}-\frac{5}{6}\right),
\end{equation}
with $M_X(\field)=\frac{1}{2}g_X\field$. The scalar one-loop contributions can be neglected 
since they are subdominant~\cite{Kierkla:2022odc}.	
	
\subsection{Thermal effective potential at leading order}
\label{sec:thermal-potential}
	
The thermal one-loop correction to the effective potential along the $\field$ direction reads
\begin{equation}
\label{eq:thermal-pot}
\VT(\field,T)=\frac{T^4}{2\pi^2}\sum_{a} 9 J_{T,b}\left(\frac{M_X(\field)}{T}\right),
\end{equation}
and the daisy correction in the Arnold-Espinosa scheme is given by~\cite{Arnold:1992} 
\begin{equation}
V_{\textrm{daisy}}(\field, T)=-\frac{3T}{12\pi}\left(M^{3}_{X,\textrm{th}}(\field,T)-M^{3}_X(\field)\right),
\end{equation}
where~\cite{Prokopec:2018}
\begin{equation}
M^2_{X, \textrm{th}}(\field,T)=M^2_X(\field)+M^2_{D,X}(T)=M^2_X(\field)+\frac{5}{6}g_X^2T^2.
\label{eq:X-mass}
\end{equation}
Since scalar loops are negligible, the only contribution affected by the daisy resummation is that of the zero Matsubara mode of the dark gauge field temporal component, which acquires a Debye mass. The full one-loop potential with daisy corrections is given by the sum of the contributions listed above
\begin{align}
\label{eq:RG-potential}
V^{\rmii {LO}}(\field,T)=\Vtree(\field)+\Vone(\field)+\VT(\field,T)+V_{\textrm{daisy}}(\field,T).
\end{align}
The label ``LO'' stands for leading order, as we will see that this potential coincides with the leading order effective potential obtained in section~\ref{subsec:EFTs}, as long as the matching is performed at leading order only (see section~\ref{sec:compare3D4D}). 

As explained in the previous section (see also ref.~\cite{Kierkla:2022odc}) in models with classical 
scale invariance we should use the RG-improved effective potential, since vastly different scales are 
present in the model. We improve the potential by replacing the field and couplings with their 
running versions as $\field \to \sqrt{Z(t)}\field$ (see the discussion in 
sec.~\ref{sec:rescaling-field-4D}), $\lambda_{\field} \to \lambda_{\field}(t)$, $g_X \to g_X(t)$, 
where $t=\log\frac{\mu}{\mu_0}$ and $\mu_0$ corresponds to the $Z$ boson mass. The $\beta$ 
functions and the anomalous dimension for the $\field$ field are listed in appendix~\ref{app:betas}.
		
Next we choose the scale $\mu$ as stated in eq.~\eqref{eq:mu-choice-1}, with no running included in the field dependent mass and $g_X$ defined at the scale $M_X$
\begin{equation}
\mu=\max\left(\frac{1}{2}g_X \field, \kappa T\right).\label{eq:mu-choice}
\end{equation}
This choice ensures that at large field values the field-dependent logarithmic term is (almost) cancelled by the field-dependent renormalisation scale,%
\footnote{
No running in the mass in eq.~\eqref{eq:mu-choice} means that we do not entirely eliminate the logarithm in the one-loop Coleman-Weinberg correction, but the remnant is tiny.
} 
while at lower field-values, for $\field<2\kappa T/g_X$, the field-dependent logarithms cancel between the zero-temperature and finite-temperature contribution, and the scale is fixed to $\kappa T$, cancelling a remaining $T$-dependent logarithm.

\subsection{Tunnelling and normalisation of the field}
\label{sec:rescaling-field-4D}
	
When defining the potentials we start from setting the values of the mass of the $X$ boson and its coupling $g_X$ at $\mu=M_X$. Following the procedure described in ref.~\cite{Kierkla:2022odc} we recover the values of all the couplings at the electroweak scale set by $M_Z$ and we define the theory at that scale. This is reflected in the choice of the reference scale $\mu_0$, see the discussion above eq.~\eqref{eq:mu-choice}. This means that the field $\field$ is defined at $\mu=M_Z$, i.e.\ at this scale it is canonically normalised (for a comprehensive discussion of the scale-dependence of scalar fields, see ref.~\cite{Andreassen:2014}). As we RG-improve the effective potential, we evolve the couplings and fields along their RG flows. This means that at other scales the field is not canonically normalised and the field is rescaled by the field renormalisation constant, $\sqrt{Z(t)}$. In the HT regime the running is frozen at $\mu=\kappa T$ and the normalisation of the field is given by $\sqrt{Z(\log\frac{\kappa T}{M_Z})}$.
	
At the same time, the usual bounce equation is derived from an action containing a canonically normalised scalar field, see eq.~\eqref{eq:BounceEOM}. This means that we cannot simply use the bounce equation with the RG-improved potential evolved down to the thermal scale. We could rederive the bounce equation in terms of the field defined at $\mu=M_Z$ but we choose to rather redefine the field for the purpose of solving the bounce equation and computing the action -- we introduce a new field that is defined at the scale $\mu=\kappa T$ and is thus canonically normalised. It is related to the old field via the rescaling $\field_{\kappa T}=\sqrt{Z(\log{\frac{\kappa T}{M_Z}})}\field_{M_Z}$. Remembering about this subtlety, we will not introduce extra subscripts on the field symbol $\field$ for simplicity of notation.%
\footnote{
As we will explain in the following section, the 3D theory is directly constructed at the thermal scale $\mu_{4D}=\kappa T$ (the values of the couplings at this scale are obtained by running from $\mu=M_Z$), thus the 3D field naturally is defined at this scale and needs no rescaling.
}
	
The factor of $\sqrt{Z}$ may seem unimportant, as $Z$ stays close to 1, as long as we are within the perturbative regime of the theory, however, it turns out that it affects the results visibly and we therefore take this $Z$ into account in the LO computations of section~\ref{sec:results}. This issue has not been appreciated in the literature, see for example ref.~\cite{Kierkla:2022odc}.

%
\section{High-temperature effective theory}
\label{sec:HT-EFT}

As mentioned in the introduction, thermal field theory suffers from poor convergence of the perturbative expansion, which can hamper the precision with which the properties of the phase transition are determined. Even though daisy resummation (see eq.~\eqref{eq:daisy}) resums a leading set of IR-sensitive diagrams, and is hence correct at $\mathcal{O}(g^3_X)$, a problem persists: parametrically large $\mathcal{O}(g^4_X)$ contributions are still missing. In particular, there is an uncancelled RG-scale dependence due to the omission of two-loop thermal masses, and furthermore additional resummations are required at the same order.%
\footnote{
These additional resummations generate contributions to the couplings, as well as momentum-dependent field normalisation contributions~\cite{Kajantie:1995dw}. 
}
The root of both problems lies in the Bose-enhancement of the low-energy modes, resulting in an enhancement of the effective parameters of these modes.

\subsection{Effective field theories}\label{subsec:EFTs}
A systematic way to deal with the above-mentioned problems, is to construct a series of effective field theories describing the thermodynamics at the different relevant energy scales. See ref.~\cite{Gould:2023ovu} for a recent discussion of the possibly relevant scales. For our purposes, it suffices to distinguish the following two energy scales (see table~\ref{tab:scales}):
\begin{table}[h]
\centering
\begin{tabular}{lcc}
\bf{Name} & \bf{Energy scale}  & \bf{Scaling of expansion parameter}   \\
Hard & $\pi T$ & $\frac{g^2}{\pi^2}$   \\
Soft & $gT$ &  $\frac{g}{\pi}$  \\
\end{tabular}
\caption{Relevant energy scales for the SU(2)cSM model (see section~\ref{sec:model}) at finite temperature. Conventionally, $g$ denotes the largest relevant coupling in the theory, in our case $g_X$.\label{tab:scales}}
\end{table}

\paragraph*{The hard scale}
For the construction of the EFTs, we make use of the partition function given by 
\begin{align}
\label{eq:partition-function}
\mathcal{Z} = \int D\Phi \; \exp\Big( -\int\limits_0^{1/T} \dd\tau \int \dd^3\mathbf{x} \; \mathcal{L}_E \Big).
\end{align}
Here, $\tau$ is the imaginary time coordinate and its periodicity is set by the reciprocal of the temperature. The functional integration $\int D\Phi$ is performed over all fields. $\mathcal{L}_E$ denotes the Euclidean space Lagrangian density. The fields can be written as a sum over momentum modes, with momenta $P = (\omega_n, \mathbf{p})$, with the Matsubara frequency $\omega_n = 2\pi n T$ for bosons and $\omega_n = (2n+1)\pi T$ for fermions. The theory described by eq.~(\ref{eq:partition-function}) contains all momentum modes, but modes with masses larger than $\pi T$ get Boltzmann-suppressed, so we see that the largest relevant energy scale in the HT regime%
\footnote{
As described in the previous sections, there are larger energy scales set by the large background field, but these scales do not admit an HT expansion.
}
in the problem is the so-called \emph{hard scale} of $\mathcal{O}(\pi T)$.

\paragraph*{The soft scale}
We can obtain the effective theory at the soft scale by formally integrating out all $n\neq 0$ momentum modes with $\omega_n \geqslant \pi T$. The resulting partition function for the theory containing only the scalar fields and gauge bosons is given formally by
\begin{align}
\label{eq:partition-EFT}
\mathcal{Z}^{\rmii{soft}}_{3} = \int D\Phi_{\omega_n = 0} \; \exp\Big( \int \dd^3 \mathbf{x} \; 
\mathcal{L}_3^{\rmii {soft}} + f_0 \Big) \equiv \int D\Phi_{\omega_n = 0} \; \exp \big( S_3^{\rmii{soft}} \big).
\end{align}
Here the path integral is over the zero modes only. The $1/T$ factor coming from the integral over 
$\tau$ is absorbed by the fields in the 3D Lagrangian such that the exponent is dimensionless, see 
the discussion below eq.~\eqref{eq:EFT-soft-tree} and e.g.\ ref.~\cite{Laine:2016hma}. The fields in 
the effective Lagrangian $\mathcal{L}_\mathrm{3}^{\rmii{soft}}$  are static and three-dimensional; 
they carry no momentum in the imaginary time direction. $f_0$ is the coefficient of the unit 
operator, related to the pressure in the symmetric phase~\cite{Braaten:1995cm}.

As we will see explicitly below, the zero modes of the temporal components of the gauge fields obtain a Debye mass from the screening by the hard modes. This mass is of the order $m_D \sim gT$, with $g$ the relevant gauge coupling. This mass scale defines the so-called \emph{soft scale}, which is the largest energy scale of the EFT constructed by integrating out the hard modes. The spatial components of the gauge fields do not get screened, and are thus lighter than the soft modes.%
\footnote{
Within the EFT, there is yet another scale deeper in the IR, often referred to as the ultrasoft scale 
$\mathcal{O}(\frac{g^2}{\pi^2} T)$. At this scale, spatial gauge field modes obtain masses due to 
non-perturbative physics.
} 
The screening of the hard scale also generates a mass $\propto g T$ for the scalar field. 

Let us now get more explicit. The action of the 3D EFT -- the action obtained after integrating out the hard modes -- separates into $S^{{\rmii {soft}}}_3 = S^{\rmii{soft, dark}}_3 + S^{\rmii{soft, SM}}_3 + \ldots$, where the ellipsis denotes the contribution from a portal that couples the two sectors together. From now on, for simplicity, we focus solely on the dark sector part within the EFT as that is what we need for the computation of the bubble nucleation rate. Yet, it is good to keep in mind that we still capture the SM contributions coming from the hard scale in the matching relations. For the dark SU(2) sector, the action reads
\begin{align}
\label{eq:EFT-action}
S^{\rmii{soft, dark}}_3 =& \;  \int \dd^3 \mathbf{x} \; \Big\{ \frac{1}{4} F^a_{ij} F^a_{ij} + (D_i \phi)^\dagger 
(D_i \phi) + \frac{1}{2} (D_i X^a_0)^2 + V_3(\phi, X^a_0) \Big\} + f_0.
\end{align}
$\phi$ now denotes the scalar field and $F^a_{ij}$ is the gauge field strength tensor of the spatial gauge field $X^a_i$ with spatial Lorentz indices $i=1,2,3$ and SU(2) isospin index $a=1,2,3$. The gauge coupling of the EFT is denoted by $g_{X,3}$. The temporal components of the gauge field, $X^a_0$, are Lorentz scalars in the EFT and transform in the adjoint representation of SU(2). The covariant derivatives for scalar doublet and triplet are 
$D_i \phi = \partial_i \phi - i g_{X,3} \frac{\tau^a}{2} X^a_i \phi$ 
and $D_i X^a_0 = \partial_i X^a_0 + g_{X,3} \epsilon^{abc} X^b_i X^c_0$, respectively.  The scalar potential reads 
\begin{align}
\label{eq:EFT-soft-tree}
V_3^{\rmii{soft, tree}}(\phi, X^a_0) &= m^2_3 \phi^\dagger \phi + \lambda_3 (\phi^\dagger \phi)^2 + \sum_{n=3}^\infty c_{2n} (\phi^\dagger \phi)^n \nonumber  \\
&+ \frac{1}{2} m^2_{D,X} X^a_0 X^a_0 + \frac{1}{4} \kappa_3 (X^a_0 X^a_0)^2 + \frac{1}{2} h_3 \phi^\dagger \phi  X^a_0 X^a_0.
\end{align}
Note that within the EFT, fields have dimension $T^{\frac{1}{2}}$, but we do not give them an explicit label ``3", and the mass terms $m_3, m_{D,X}$  and the couplings $\lambda_3,\kappa_3,h_3$ have dimension $T$. We include marginal operators for the doublet $\phi$ with coupling constants $c_n$, which correspond to the terms containing the $\zeta$-terms in eq.~(\ref{eq:T-master-high-T}). We use these marginal operators only as an indicator of a breakdown of the HT expansion: when their effect becomes non-negligible at low temperature, or more importantly, at large field values, the HT expansion starts to break down.

The parameters of the 3D theory are obtained by a matching procedure, as is common in the 
construction of any EFT. We use {\tt DRalgo}~\cite{Ekstedt:2022bff} for the determination of the 
parameters of the soft scale EFT, and list the result in appendix~\ref{sec:Matching}. We highlight 
that the momentum-dependent field normalisation contributions of the hard modes are absorbed 
into the parameters of the EFT for all fields, rather than including $Z$-factors in the kinetic terms of 
the soft scale EFT action~\cite{Kajantie:1995dw}. For illustration of the matching procedure, see 
e.g.\ refs.~\cite{Kajantie:1995dw,Schicho:2021gca}, or appendix B.1 in ref.~\cite{Croon:2020cgk}. 
We emphasise that the construction of the 3D EFT is performed in the symmetric phase, and relies 
on the high-temperature expansion for the matching, which assumes that $m/T \ll 1$ for all the 
fields. In conformal models all fields are massless in the symmetric phase, yet deep in the broken 
phase the assumption about the HT expansion is no longer valid, as discussed in 
section~\ref{sec:HT-LT-regimes}.

We now assign a background field to the scalar field
\begin{equation}
\phi = \frac{1}{\sqrt 2} \begin{pmatrix} 0 \\ v_3 \end{pmatrix},
\end{equation}
resulting in masses for the spatial and temporal gauge bosons respectively 
\begin{equation}
m_{X,3}^2 = \frac 1 4 g_{X,3}^2 v_3^2, \qquad m^2_{X_{0},3} = m_{D,X}^2 + \frac 1 2 h_3 v_3^2. \label{eq:mass-gauge}
\end{equation}
The tree-level potential for the field $v_3$ at the soft scale is then given by
\begin{equation}
\label{Veff:softTree}
V_3^{\rmii{soft, tree}}(v_3) = \frac 1 2 m_3^2 v_3^2 + \frac 1 4 \lambda_3 v_3^4 + 
\sum_{n=3}^\infty \frac{1}{2^n} c_{2n} v_3^{2n}.
\end{equation}
Note that as explained above, the marginal operators with coefficients $c_{2n}$ are suppressed in 
the HT expansion and hence of higher order, and included only for inspecting  departure from the 
HT regime at large field values. 
In order to obtain a cubic term, we have to integrate out the gauge field modes, resulting in a new 
EFT expansion for the effective potential and the effective action. This is a generic feature of models 
where the tree-level potential does not include a barrier\footnote{
	As $\lambda_3$ is negative, strictly speaking, we already have a barrier in the potential of 
	eq.~(\ref{Veff:softTree}), but the correction of the cubic term was observed to be large 
	\cite{Kierkla:2022odc}.
} required for first-order phase transition~\cite{Ekstedt:2020abj,Gould:2021ccf, 
Ekstedt:2022zro,Lofgren:2023sep,Gould:2023ovu}. Since all the masses in this theory are formally 
soft, it is not immediately obvious that we can integrate out the gauge modes. We return to the issue 
of mass hierarchies in section~\ref{sec:scale-shifters}.

Integrating out the gauge field modes at one-loop order results, together with the tree-level term of the soft scale EFT, in the leading order contribution of the final EFT expansion%
\footnote{
Note that this potential should not be confused with the $V^{\rmii{LO}}$ potential of eq.~\eqref{eq:RG-potential}, as $V_3^{\rmii {EFT, LO}}$ uses the full matching relations, whereas $V^{\rmii{LO}}$ does not (see section~\ref{sec:compare3D4D}). 
}
\begin{align}
\label{eq:Veff-EFT-LO}
V_3^{\rmii {EFT, LO}} =&\ \frac 1 2 m_3^2 v_3^2 + \frac 1 4 \lambda_3 v_3^4 + 
\sum_{n=3}^\infty \frac{1}{2^n} c_{2n} v_3^{2n} \nonumber \\
&- \frac{1}{12\pi} 
\Big( 6 (m^2_{X,3})^{\frac{3}{2}} + 3 (m^2_{X_{0},3})^{\frac{3}{2}} \Big) 
+ \frac{1}{12\pi} \Big( 3 (m^2_{D,X})^{\frac{3}{2}} \Big).
\end{align}
Here the last, field-independent, term has been added to normalise the potential to zero for a vanishing field. The validity of this EFT expansion will be further discussed in sections~\ref{sec:validityEFT} and~\ref{sec:scale-shifters}, and higher-order corrections are determined in section~\ref{sec:higher-orders}.

\subsection{Relation to one-loop thermal effective potential}\label{sec:compare3D4D}

One can confirm straightforwardly that eq.~\eqref{eq:RG-potential} in the HT expansion is reproduced exactly by eq.~\eqref{eq:Veff-EFT-LO} when the matching relations (c.f.\ appendix~\ref{sec:Matching}) are truncated as follows
\begin{align}
\lambda_{3} &= T \Big( \lambda_{\field} + \frac{1}{(4\pi)^2} \frac{3}{8} g^4_X (1 - \frac{3}{2} L_b) \Big), \label{eq:lambda3-LO}\\
g_{X,3}^2 &= g_X^2 T, \label{eq:gX3-LO}\\
h_3 &= \frac{1}{2} g^2_X T, \\ 
\label{eq:mDsq-lo}
m_{D,X}^2 &= \frac{5}{6} g^2_X T^2, \\
\label{eq:m3sq-lo}
m^2_{3} &= \frac{3}{16} g^2_X   T^2,
\end{align}
and taking into account dimensional scalings $v_3=\field/\sqrt{T}$ and $V^{\rmii {LO}}(\field, T) = T V_3^{\rmii {EFT, LO}} $. Note that the Coleman-Weinberg term is captured by the $L_b$ term in $\lambda_3$, where $L_b = 2 \gamma_E - 2\log{[4\pi]} + \log{\left[\frac{\mu_4^2}{T^2} \right]}$ (see eq.~\eqref{eq:LbLf}). We note that while we simply took eq.~\eqref{eq:daisy} for leading daisy resummation from the literature, in eq.~\eqref{eq:Veff-EFT-LO} we actually derived it in the 3D EFT approach and we can clearly see the physics behind it: this term originates from the fact that hard modes screen the soft zero mode, and once this hard scale screening is accounted for by the soft mode mass at one-loop order, the EFT automatically resums this contribution to all orders. In the EFT, a one-loop computation with the resummed propagator is easy and, furthermore, two-loop diagrams are also straightforward, and we exploit this fully in the next section.

\subsection{Higher-order corrections within the EFT}
\label{sec:higher-orders}

Now that we have illustrated what is behind the EFT approach, let us highlight what happens when we do not truncate the matching relations: in this case, we account for thermal corrections of the hard scale screening in not just the mass, but also in all couplings, at one-loop order. This includes momentum-dependent contributions, i.e.\ we account for field renormalisation factors, see ref.~\cite{Kajantie:1995dw}. Even more importantly, we include two-loop corrections in the thermal mass, and together with field renormalisation factors this guarantees that 3D EFT parameters are renormalisation scale invariant to the order we compute. 
 
The construction of the NLO contribution to the potential follows the prescription of refs.~\cite{Gould:2021ccf, Hirvonen:2021zej, Hirvonen:2022jba}. We include two-loop contributions of the soft modes, which yield the NLO correction to the effective potential~\cite{Ekstedt:2022bff,Gould:2023ovu}
\begin{align}\label{eq:VeftNLO}
V_3^{\rmii{EFT,NLO}} &= 
\frac{1}{(4\pi)^2} \bigg\{ \frac{3}{64} g^2_{X,3} \bigg( 56 m^2_{X,3} \Big(1-3\ln(3) \Big) + g^2_{X,3} v^2_{3} \Big(2-\ln(256) \Big) \nonumber \\
& + 2 (80 m^2_{X,3} - 3 g^2_{X,3} v^2_{3}) \ln\Big( \frac{\mu_3}{2 m_{X,3}} \Big) \bigg)  \bigg\} \nonumber \\
& + \frac{1}{(4\pi)^2} \bigg\{ \frac{3}{4} g^2_{X,3} (6 m^2_{X_0,3} + 4 m_{X,3} m_{X_0,3} - m^2_{X,3} ) 
+ \frac{15}{4}  \kappa_3 m^2_{X_0,3} \nonumber \\
& - \frac{3}{8} h^2_{3} v^2_3 \bigg( 1 + 2 \ln\Big( \frac{\mu_3}{2 m_{X_0,3}} \Big) \bigg) - \frac{3}{2} g^2_{X,3} (m^2_{X,3} - 4 m^2_{X_0,3}) \ln\Big( \frac{\mu_3}{2 m_{X_0,3} + m_{X,3} } \Big)   \bigg\}  \nonumber \\
&-\frac{1}{(4\pi)^2} \left\{ \frac{15}{4} m_{D,X}^2 \kappa_3 + \frac{3}{2} g_{X,3}^2 m_{D,X}^2\left(3 + 4 \ln{\left(\frac{\mu_3}{2 m_{D,X}} \right)}\right) \right\} .
\end{align}
The last line is independent of the field value, and ensures that the potential is zero at the origin. We will denote the sum of the LO and NLO potential as
\begin{equation}
\label{eq:VEFT}
V_3^{\rmii{EFT}}(v_3) = V_3^{\rmii{EFT, LO}}(v_3) + V_3^{\rmii{EFT, NLO}}(v_3).
\end{equation}

In addition to the corrections to the effective potential, integrating out the gauge modes also results in a field normalisation term for the effective action (note that we normalise the potential such that it is zero at the origin, which also implies discarding $f_0$ in eq.~\eqref{eq:EFT-action})
\begin{equation}
S_{\rm 3}^{\rmii{EFT}} =4\pi\int \dd  r\   r^2\left( \frac{1}{2} Z_3(v_3)(\partial_i v_3)^2 + V_3^{\rmii{EFT}}(v_3) \right),
\end{equation}
with $Z_3(v_3)$ given by (see also refs.~\cite{Moore:2000jw, Ekstedt:2022ceo, Gould:2022ran})
\begin{equation}
\label{eq:Z3}
Z_3(v_3) = 1 + Z^{\rmii{NLO}}_3(v_{3}) = 1-\frac{11 g_{X,3}}{16\pi v_3} + \frac{h_3^2 v_3^2}{64 \pi m_{X_0,3}^3}.
\end{equation}
Within the EFT, the term with the $Z_3$-factor is an effective derivative operator generated by the gauge modes. The role of this $Z_3$-factor within the EFT is different from field renormalisation in the parent theory, as it is not related to UV running.

Let us pay close attention to the soft 3D EFT RG-scale, $\mu_3$. We can split the action into two parts
$S_{\rm 3}^{\rmii{EFT}} = S_{\rm 3}^{\rmii{EFT,LO}} + S_{\rm 3}^{\rmii{EFT,NLO}}$, where
\begin{align}
S_{\rm 3}^{\rmii{EFT,LO}} &= 4\pi\int \dd {\boldmath r}\ {\boldmath r}^2\left(\frac{(\partial_i v_{3})^2}{2} + V_3^{\rmii{EFT,LO}}(v_{3})\right), \\
S_{\rm 3}^{\rmii{EFT,NLO}} &= 4\pi\int \dd {\boldmath r}\ {\boldmath r}^2\left(\frac{1}{2} Z^{\rmii{NLO}}_3(v_{3})(\partial_i v_3)^2 + V_3^{\rmii{EFT,NLO}}(v_{3})\right),\label{eq:SbounceNLO}
\end{align}
where NLO is suppressed compared to LO by the soft expansion parameter, formally $\mathcal{O}(\frac{g_X}{\pi})$. The LO action depends explicitly on the 3D EFT mass parameter, $m_3$, which runs (see eq.~(\ref{eq:3DMass})).%
\footnote{
Note that other parameters of the EFT do not run with the soft 3D renormalisation scale.
} 
The NLO part of the action is independent of the scalar mass, yet contains explicit logarithms of the RG-scale.  In ref.~\cite{Hirvonen:2021zej} it has been shown that the action is RG invariant at the order considered ($\mathcal O (g_X^4)$). For completeness, we reproduce this argument here. We simply have
\begin{align}
\label{eq:3D-scale-invariance}
\mu_3 \frac{\dd S_3^{\rmii{EFT}} }{\dd\mu_3} = \beta_{m^2_3} \frac{\partial S_3^{\rmii{EFT,LO}} }{\partial m_3^2} + \mu_3 \frac{\partial S_3^{\rmii{EFT,NLO}} }{\partial \mu_3} +\mathcal{O}(g_X^{5})= \mathcal{O}(g_X^{5}),
\end{align}
as the running of the scalar mass is governed by the beta function
\begin{align}
\beta_{m^2_3} \equiv \mu_3 \frac{d m^2_3}{d\mu_3} = \frac{1}{(4\pi)^2} (-39 g^4_{X,3} - 96 g^2_{X,3} h_3 + 24 h^2_3),
\end{align}
which exactly cancels the $\mu_3$-dependent terms in $V_3^{\rmii{EFT,NLO}}$.

In our computation, we use the freedom to choose the RG scale and use $\mu_3 = \kappa_{\rmii{RG}} m_{X,3}$, where the coefficient $\kappa_{\rmii{RG}}$ is of the order unity. Should one aim to optimise the choice of RG-scale, one could follow e.g.\ refs.~\cite{Farakos:1994kx, Ghisoiu:2015uza}.

\subsection{Thermal parameters at NLO}
\label{sec:nucleation-in-EFT}

To compute the exponential part of the nucleation rate in the EFT, we follow refs.~\cite{Gould:2021ccf, Hirvonen:2021zej}. We use a method based on the strict expansion of the action in order to obtain results that are gauge invariant. 

In the strict expansion, the critical bubble configuration $v_{3,B}$ is formally expanded as $v_{3,B} = v_{3,b} + \mathcal{O}(\frac{g}{\pi} T^{\frac12})$ and the LO bounce $v_{3,b}$ is found only using the LO effective potential:
\begin{align}
\Box v_{3,b}(r) = \frac{\partial}{\partial v_3} V_3^{\rmii{EFT,LO}}(v_{3,b}), 
\end{align}
with boundary conditions $\partial_r v_{3,b}(0)=0$ and $v_{3,b}(r\to\infty)=0$. 

The NLO action is then simply evaluated at the leading order bounce, and higher-order corrections to the bounce result in contributions that are formally beyond our accuracy goal~\cite{Hirvonen:2021zej, Ekstedt:2022ceo}. It has been shown in refs.~\cite{Lofgren:2021ogg,Hirvonen:2021zej} that despite the singularity in the $Z$-factor at zero field value, its contribution to the action is finite.

The bounce solution depends on the scalar mass and hence inherits RG-scale dependence through it. Note, however, that the implicit running of the bounce does not contribute at the order considered, because the LO action is extremised by the bounce, i.e. when applying the chain rule in eq.~\eqref{eq:3D-scale-invariance}, we see that the term $\mu_3 \frac{\partial }{\partial \phi_{b}} S_3^{\rmii{EFT,LO}}  \frac{d \phi_b}{d\mu_3}$ vanishes. 

For illustration, let us use a simplified expression for the nucleation condition, i.e.\ that the nucleation rate equals the Hubble parameter, $H$
\begin{equation}
\frac{A(T_n) e^{-S(T_n)}}{H^4} \approx 1 \quad \rightarrow \quad S(T_n) = 4\log \left[\frac{A(T_n)^{1/4}}{H(T_n)} \right]. 
\end{equation}
Here we will use
\begin{equation}
A(T) = T^4, \qquad H(T)^2 = \frac{1}{3 M_{\rm pl}^2} \left(\frac{T^4}{\xi_g^2} + \Delta V \right),
\end{equation}
i.e.\ we estimate the prefactor simply using dimensional analysis since we do not compute this contribution properly, but only assume it is suppressed compared to the exponential part. $M_{\rm pl}$ is the reduced Planck mass, and $\xi_g = \sqrt{30/(g_* \pi^2)}$ with $g_*$ the number of degrees of freedom in the plasma. For convenience, we define the shorthand notation $S_0 \equiv S_3^{\rmii{EFT,LO}}[v_{3,b}]$ and $S_2 \equiv S_3^{\rmii{EFT,NLO}}[v_{3,b}]$. We can determine the nucleation temperature by the \textit{mixed method}%
\footnote{
In our numerics, we do not encounter the problems reported in ref.~\cite{Gould:2023ovu}. There, the use of the mixed method can lead to spurious IR divergences in the determination of the critical temperature, since the LO minimum for the strict expansion of the effective potential does not necessarily exist for wider temperature ranges around the leading order critical temperature. For the nucleation rate, we do not encounter this problem, as the LO bounce exists over a wide temperature range.    
} 
(c.f.\ ref.~\cite{Gould:2023ovu}) where we compute the action in the strict expansion, yet we directly solve for the nucleation temperature from the condition 
\begin{align}
S_0(T_n) + S_2(T_n) = 2 \log{\left[ \frac{3 M_{\rm pl}^2 T_{n}^2 }{T_{n}^4/\xi_g^2 + \Delta V}\right]}.
\end{align}
This method is not only gauge invariant, but also invariant with respect to the soft 3D EFT renormalisation scale, since the sum $S_0 + S_2$ is invariant. In this sense, only using the sum $S_0 + S_2 $ describes the full NLO soft scale corrections. In our numerical analysis we apply this method to compute the nucleation and percolation temperatures, i.e.\ we evaluate the NLO action at the LO bounce solution and use the resulting $S_0(T)+S_2(T)$ with the standard formulas for nucleation and percolation temperatures, following the approach described in ref.~\cite{Kierkla:2022odc}.

In analogy to ref.~\cite{Gould:2023ovu}, we could use the strict expansion for the nucleation temperature as well, by formally expanding in $\delta$ (which is set to 1 at the end of the computation): $S = S_0 + \delta^2 S_2 + \mathcal{O}(\delta^3)$ and $A = \delta^3 A_3 + \mathcal{O}(\delta^4 T^4)$~\cite{Gould:2021ccf}, and furthermore expand $T_n = T_{n,0} + \delta T_{n,1} + \delta^2 T_{n,2} +  \mathcal{O}(\delta^3)$. Here $\delta$ denotes suppression with respect to $S_0$. We can then expand the simplified nucleation condition, and by equating the first two non-vanishing orders we find that $T_{n,0}$ is given by
\begin{align}
S_0(T_{n,0})  = 2\log \left[\frac{\sqrt{A_3(T_{n,0})} }{H^2(T_{n,0})} \right],
\end{align}  
and the first non-zero correction to it reads
\begin{equation}
T_{n,2} = \frac{S_2}{\frac{A_3'}{A_3} - S_0' - 4 \frac{H'}{H}} = -\frac{S_2 T_{n,0}(T_{n,0}^4/\xi_g^2 + \Delta V)}{4(T_{n,0}^4/\xi_g^2-\Delta V  ) + T_{n,0}(T_{n,0}^4/\xi_g^2 + \Delta V)S_0'},
\end{equation}
where all quantities are evaluated at $T_{n,0}$ and in the second expression we assumed again $A_3 = T^4$. However, we do not find this expansion useful as the sum $T_{n,0} + T_{n,2}$ appears not to be scale invariant, and neither are the individual terms and we therefore choose to use the mixed method.%
\footnote{
We could choose the 3D scale so that NLO action at $T_{n,0}$ is negligible, and as explained in the main body, this effectively resums the information of the NLO action to the LO action, and hence into $T_{n,0}$ itself. For such construction, we find agreement with the mixed method. 
Yet, this strict expansion does not immediately generalise in the case of the full nucleation condition or the determination of the percolation temperature.
}

\subsection{Power counting and validity of the EFT}
\label{sec:validityEFT}

Now that we have given our expressions for the LO and NLO effective potential, as well as the approach to determine the nucleation rate and temperature, let us understand better in which sense eq.~\eqref{eq:Veff-EFT-LO} describes the LO behaviour, and what kind of power counting this implies for perturbation theory.

For starters, let us consider the scaling of different contributions to the potential in the presence of a radiatively generated barrier. The existence of the barrier requires~\cite{Arnold:1992}
\begin{align}
\label{eq:scaling}
m_3^2 v_3^2 \sim \lambda_3 v_3^4 \sim \frac{g^3_{X,3} v^3_3}{\pi}, 
\end{align}
i.e.\ all terms in the potential are of the same order. As discussed in detail in ref.~\cite{Gould:2023ovu}, one possible realisation of this occurs in the temperature regime where the scalar mass parameter is parametrically lighter than soft, in particular $m_3^2 \sim \Big(\frac{g^{\frac{3}{2}}}{\sqrt{\pi}} T \Big)^2 \ll (g T)^2$, which further implies $\lambda_{3} \sim \frac{g^3_{X}}{\pi} T$ and $v_3 \sim \sqrt{T}$. In such case, the gauge field modes are soft $m_{X,3}, m_{X_{0},3} \sim g T$, and eq.~\eqref{eq:Veff-EFT-LO} is interpreted as the construction of the \textit{supersoft} EFT below the soft scale. In ref.~\cite{Gould:2023ovu} such effective description exists, since the mass parameter has the schematic form $m^2_3 = m^2_0 + C g^2 T^2$, where $C$ is a positive constant and $m^2_0$ is a negative zero temperature mass parameter. The partial cancellation of these two terms, which are individually of order $(g T)^2$, makes it possible that the effective scalar mass is parametrically lighter than soft in some temperature range. The effective description for the phase transition is constructed in such a temperature window, and furthermore the background field is assumed to scale as $v_3 \sim \sqrt T$ in the broken phase, such that the gauge modes are soft. The potential itself follows the scaling 
\begin{align}
V_3^{\rmii {EFT, LO}} \sim \frac{g^3}{\pi} T^3 \Big( 1 + \frac{g^3}{\pi^3} + \mathcal{O}(\frac{g^5}{\pi^5}) \Big),
\end{align}
where corrections are due to marginal operators with $c_{2n} \sim 
\frac{g^{2n}}{\pi^{2(n-1)}}$~\cite{Croon:2020cgk}. At parametrically smaller field values $v_3 \sim 
\sqrt{g T}$ the effective description is compromised. Indeed, in the terminology used in 
ref.~\cite{Gould:2021ccf} gauge field modes are \textit{scale-shifters}: at field values close to the 
symmetric phase, in the bubble tail, the effective description based on a derivative expansion of the 
effective action alone fails, yet can be used at the bulk of the bubble to compute the nucleation 
rate~\cite{Ekstedt:2021kyx,Gould:2021ccf,Hirvonen:2021zej,Ekstedt:2022tqk,Ekstedt:2022ceo}. We 
will return to this issue below.  

In our case of a dimensionally transmutated theory, the previous discussion becomes more subtle and needs to be modified, since the scalar mass parameter cannot be lighter than soft $m^2_3 \sim (g T)^2$, since there is no zero temperature mass $m^2_0$. In this case, eq.~\eqref{eq:scaling} implies $\lambda_{3} \sim \frac{g^4_{X}}{\pi^2} T$ and $v_3 \sim \frac{\pi}{g} \sqrt{T}$.%
\footnote{
Formally, this scaling for the field was derived close to the root of the potential, behind the potential 
barrier. We have also checked that this scaling holds approximately at the escape point of the 
bounce trajectory.
}
Such a huge value of the background field pushes the gauge field modes \textit{formally} to the hard scale $m_{X,3}, m_{X_{0},3} \sim \pi T$ and furthermore signals a possible breakdown of the HT expansion, as the scaling of the potential becomes
\begin{align}
V_3^{\rmii {EFT, LO}} \sim \pi^2 T^3,
\end{align}
and marginal operators are no longer strongly suppressed, but could contribute at leading order. 

Does this mean that the effective description based on the effective potential that we derived 
cannot be used after all? We argue that this is not the case, for the following reasons: Even though 
the aforementioned formal power countings for masses and the background field can bring clarity 
about how to organise the perturbative expansion, it is not clear how strictly they should be 
followed. As scale-shifters, the gauge field modes vary over multiple mass scales, from the hard to 
the soft to eventually the non-perturbative ultrasoft scale ($\frac{g^2}{\pi^2} T$). As the bounce 
solution interpolates between the two phases, the bubble nucleation rate obtains contributions from 
different scales. Intuitively, the EFT description could capture most of the effects reliably, provided 
that         
\begin{itemize}
\item[1)] Non-zero Matsubara modes are much heavier than zero modes.
\item[2)] Gauge field zero modes are much heavier than the nucleating scalar field zero mode.
\end{itemize}
In the following subsection we study the mass hierarchies in our model in more detail and explain 
how we treat the scale-shifters.

\subsection{Mass hierarchies and scale-shifters}\label{sec:scale-shifters}

\begin{figure}[!ht] 
\centering
\includegraphics[width=0.5\textwidth]{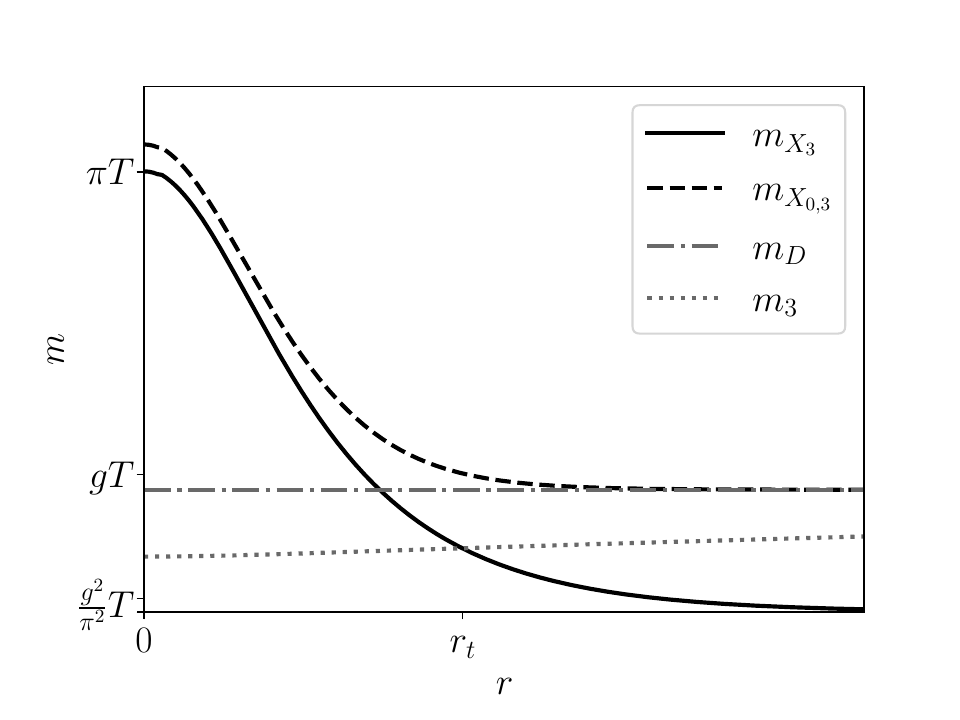}
\caption{
Schematic mass hierarchies we encounter along the bounce solution as a function of the radial coordinate.
}
\label{fig:mass-hierarchy}
\end{figure}

Let us study the typical mass hierarchies of the problem, which we encounter when using the potential of eq.~\eqref{eq:Veff-EFT-LO} to find the bounce solution using the schematic figure~\ref{fig:mass-hierarchy}.
In this figure, the masses of the gauge field modes $m_{X,3}, m_{X_{0},3}$ are given as a function of the critical bubble radial coordinate $r$, together with soft mass parameters $m_D$ and $m_3$.%
\footnote{
Note that while $m_D$ is a constant, the scalar mass parameter $m_3$ has a mild dependence on the radial coordinate: this dependence arises since the two-loop thermal mass is a function of the 3D scale $\mu_3$ that we fix to $m_{X,3}$ which varies along the bounce, c.f.\ section~\ref{sec:higher-orders}. Note also that figure~\ref{fig:mass-hierarchy} is not invariant with respect to $\mu_3$, as the bounce itself is not.
}  
Indeed, both $m_D$ and $m_3$ are parametrically of the order $g T$, i.e. soft, yet note that due to group theory factors in the LO dimensional reduction matching relations they differ approximately by a factor 2, c.f.~eqs.~\eqref{eq:mDsq-lo},~\eqref{eq:m3sq-lo}.  Close to the center of the bubble at small $r$, $m_{X,3}, m_{X_{0},3}$ indeed become very large, yet they are still below the lightest bosonic non-zero Matsubara mode with mass $2\pi T$ at the escape point. In this case, one needs to be very cautious with the HT expansion. However, we demonstrate in sec.~\ref{sec:PlotVs} that indeed up to the escape point, the HT expansion converges well and as a consequence also marginal operators are suppressed, providing support that the EFT picture is reliable in the small $r$ regime. 

On the other hand, we know that the EFT picture fails at the \textit{bubble tail} $r>r_t$, where $r_t$ is defined at the radial distance where the spatial gauge mode mass becomes comparable with the nucleating scalar mass, suggesting that it is not possible to integrate out the gauge modes. Therefore, we can trust the EFT that we constructed for field values above $v_3>v_{3,b}(r_t)$.

Finally, we emphasise the following: as long as points 1) and 2) stated at the end of 
section~\ref{sec:validityEFT} are valid, the higher-order corrections to eq.~\eqref{eq:Veff-EFT-LO} 
given in eq.~\eqref{eq:VeftNLO} are of the same form regardless of the assumed formal power 
counting for the gauge field modes and the background field. Indeed, the EFT expansion has the 
same functional form as long as $2\pi T \gg m_{X,3}, m_{X_{0},3} \gg m_3 \gg \frac{g^2}{\pi^2} T $, 
yet we do not need to fix the formal power counting for these in-between scales between the hard 
and ultrasoft scales, and indeed we cannot, since the gauge field modes are 
scale-shifters~\cite{Gould:2021ccf}.

In ref.~\cite{Gould:2021ccf} it is explained in detail how to treat the nucleation EFT construction with a scale shifter. In essence, the one-loop contribution from the gauge fields is still resummed to the LO effective potential to provide a barrier, and this affects the LO bounce solution. For the action however, their contribution is computed without derivative expansion, i.e.\ they contribute in the prefactor in analogy to the soft scalar modes. Then, one needs to subtract this gauge field contribution from the exponential part of the rate, to avoid double counting. In our analysis, we do not compute the prefactor, and hence we stick to the procedure described earlier in section~\ref{sec:nucleation-in-EFT}. We check the accuracy of this approach by estimating the contribution of the tail of the bounce, where the assumed mass hierarchies are violated, for details see section~\ref{sec:tail}.

%
\section{Numerical results}
\label{sec:results}

In this section we present the results of a scan of the entire allowed parameter space using the mixed method described in section~\ref{sec:nucleation-in-EFT} 
for computing the action with the NLO effective potential 
of eq.~\eqref{eq:VEFT} and the $Z_3$  of eq.~\eqref{eq:Z3}. We use this action to compute the nucleation and percolation 
temperatures, as well as the normalised radius of bubbles at the moment of collision 
($R_*H_*$) and the efficiency factor for bubble collisions following the procedures 
described in ref.~\cite{Kierkla:2022odc}. We also evaluate the expected observability of 
the signals in terms of the signal-to-noise (SNR) ratio, using the spectra for supercooled phase transitions from ref.~\cite{Lewicki:2022pdb}. Moreover, we compare the NLO 
predictions to the LO ones, obtained from eq.~\eqref{eq:RG-potential}, updating them with respect to ref.~\cite{Kierkla:2022odc} by 
including the thermal cutoff on the running also in the zero-$T$ part of the effective 
potential (see section~\ref{sec:HT-LT-regimes} for details) and by redefining the field to 
be canonically normalised at the thermal scale (see section~\ref{sec:rescaling-field-4D}).

\subsection{Effective potential at LO and NLO}
\label{sec:PlotVs}

\begin{figure}[!ht] 
\centering
\includegraphics[height=.31\textwidth]{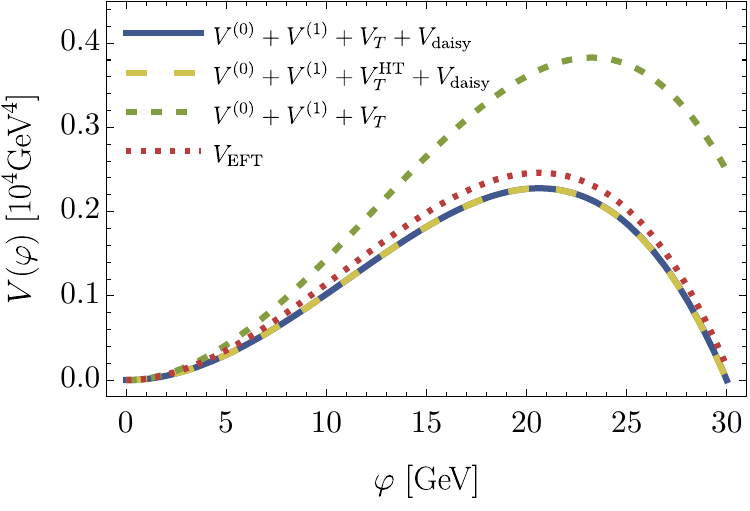}\hspace{2pt}
\includegraphics[height=.31\textwidth]{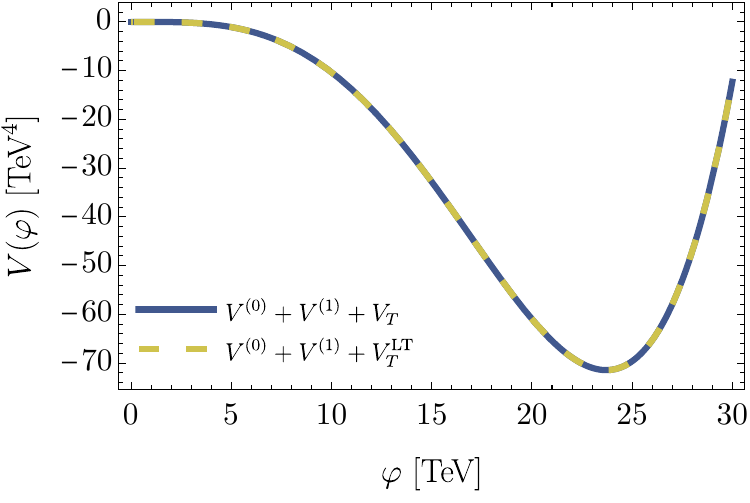}
\caption{Comparison of different approximations to the effective potential at low field values, around the barrier (left) and at large field values, around the minimum (right) at  $T_n=14.59$~GeV, for $g_X=0.8$, $M_X=10$\ TeV.}
\label{fig:comparison-potentials}
\end{figure}

Let us start with comparing potentials computed with different approximations, all evaluated at $T_n$ computed from the NLO potential. 
In the left panel of figure~\ref{fig:comparison-potentials} we focus on the low-field value or 
high-temperature regime, the range for the plot is chosen such that the barrier is well visible. The blue solid 
line represents the full one-loop potential of eq.~\eqref{eq:Veff-1loop-daisy} (see also eq.~\eqref{eq:RG-potential}), with the daisy term included. 
It agrees very well
with the high-temperature approximation of eq.~\eqref{eq:HT-potential} (long-dashed light green). 
The NLO potential computed within the EFT of eq.~\eqref{eq:VEFT} (dotted red line) differs from the LO result mildly, while exclusion of the daisy term (dashed green line) modifies the result significantly, which indicates that the daisy diagrams are indeed very relevant for the shape of the potential.%
\footnote{
Note that e.g.\ in refs.~\cite{Salvio:2023qgb,Salvio:2023ynn} it was argued that for supercooled phase transitions the high-temperature resummations are not relevant. The results presented here can be treated as an explicit check of this claim, and we reach a different conclusion.
} 
The right panel of figure~\ref{fig:comparison-potentials} shows the large-field or low-temperature behaviour of the effective potential. The full one-loop LO potential of eq.~\eqref{eq:Veff-1loop} (solid blue line) is closely approximated by the low-temperature approximation with only the first term in the sum in eq.~\eqref{eq:low-T} (long-dashed light green line).

\begin{figure}[!ht] 
\centering
\includegraphics[height=.31\textwidth]{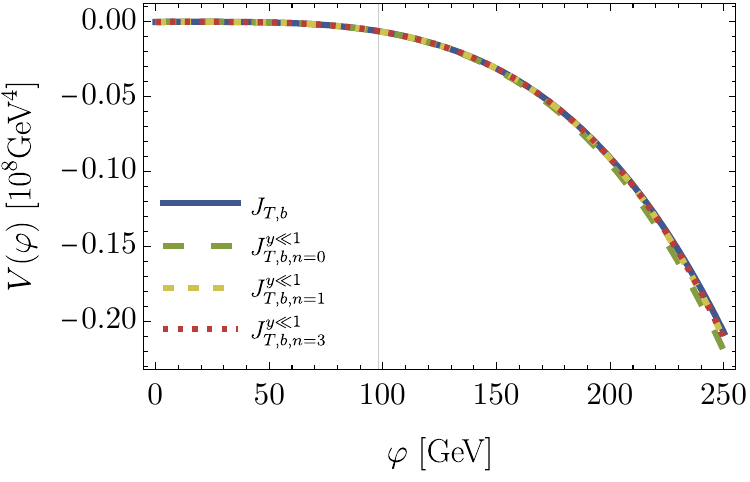}\hspace{2pt}
\includegraphics[height=.31\textwidth]{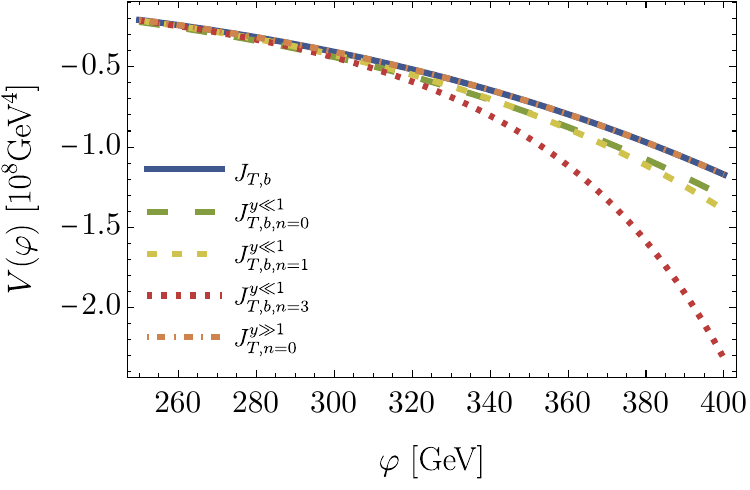}
\caption{Illustration of the validity of the high-temperature expansion at $T_n=14.59$\ GeV for $g_X=0.8$, $M_X=10$\ TeV, for field-values in 
the vicinity of the escape point (vertical grey line) and beyond (left panel), and for larger field values (right panel).}
\label{fig:HT-validity}
\end{figure}

From figure~\ref{fig:comparison-potentials} we learn that the HT expansion works perfectly around 
the barrier but we should ask whether it is valid all the way up to the escape point for the bounce trajectory and beyond. 
This is shown in figure~\ref{fig:HT-validity}. The left panel presents the potential evaluated using different approximations around 
the escape point, which is marked by the vertical grey line (obtained with the NLO action). 
The solid blue line shows the full one-loop LO effective potential, while the dashed lines indicate the usage of HT expansion of the 
thermal function, eq.~\eqref{eq:T-master-high-T} (see also eq.~\eqref{eq:Veff-EFT-LO}): long-dashed, green is the first approximation 
without the sum containing the $\zeta$-terms in the second line, short-dashed light green includes the first term in the sum, while the dotted red line includes the first 
three terms from the sum. It is clear that all the approximations agree very well in the vicinity of the barrier and beyond, only on the verge of 
the displayed region small differences between the curves can be noticed, as the first approximation to the potential deviates slightly from the full solution. 
For larger field values, shown in the right panel of figure~\ref{fig:HT-validity}, the HT expansion is quickly invalidated which is clear from the fact 
that the approximations with more terms from the HT expansion included behave worse than the ones with fewer terms. At the same time, the LT approximation 
(dot-dased orange curve) works like a charm (it overlaps with the solid blue line representing the full potential). 

This is an explicit confirmation of our earlier claims that the high-temperature expansion can be used for the field values relevant to the tunnelling, while the low- or zero-temperature potential describes accurately physics associated with the minimum of the potential.

\subsection{Phase transition and GW signal}\label{sec:GWs}

For strongly supercooled phase transitions the only parameters that are relevant for the determination of the 
gravitational wave spectrum are the length or time scale of the transition evaluated at the percolation temperature 
and the reheating temperature.

The process of reheating is controlled by the decay rate of the scalar field, which measures its ability to transfer the energy to the SM plasma. When it is larger than the Hubble parameter, reheating can be considered instantaneous. Then the Universe reheats to the temperature at which thermal inflation started,
which is controlled by the potential in the low temperature/large field regime. In ref.~\cite{Kierkla:2022odc} it was shown that this is the case for most of the parameter space of the SU(2)cSM model. Only  for low $g_X$ and high $M_X$, the decay rate of the scalar field becomes smaller than the Hubble parameter. In this case the Universe cannot reheat immediately after the phase transition. This results in a period of matter domination when the scalar field oscillates around the minimum until the decay rate becomes large enough to transfer the energy to the SM sector~\cite{Barenboim:2016mjm,Allahverdi:2020bys, Ellis:2020}. In this scenario of inefficient reheating, the final temperature is lower and the GW spectrum is modified by the modified expansion history. In ref.~\cite{Kierkla:2022odc} the region of inefficient reheating was excluded by the percolation criterion. As we will see, this region opens up by including the NLO effects, however,   it is still very small and we will not analyse it in detail since it is beyond the main focus of the present work. Thus, we will assume that the reheating temperature is given by $T_V$ and will not be changed compared to the analysis of ref.~\cite{Kierkla:2022odc}. Therefore, we will not show the results for $T_r$ here.

\begin{figure}[!ht] 
\centering
\includegraphics[width=.4\textwidth]{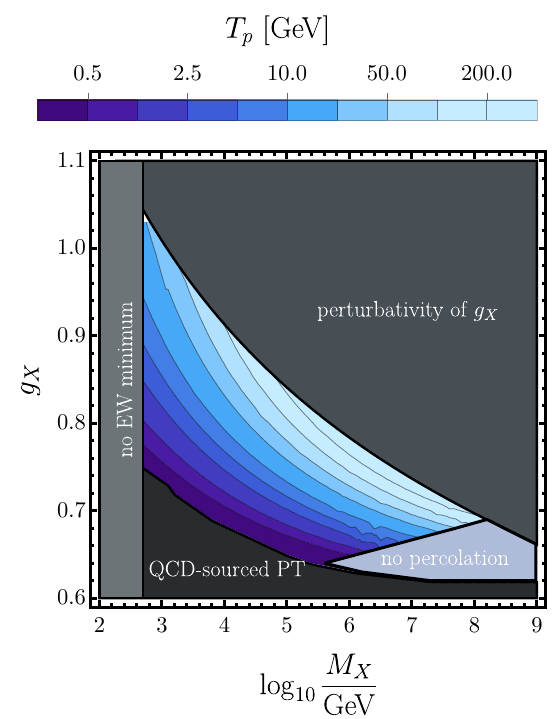}\hspace{2pt}
\includegraphics[width=.4\textwidth]{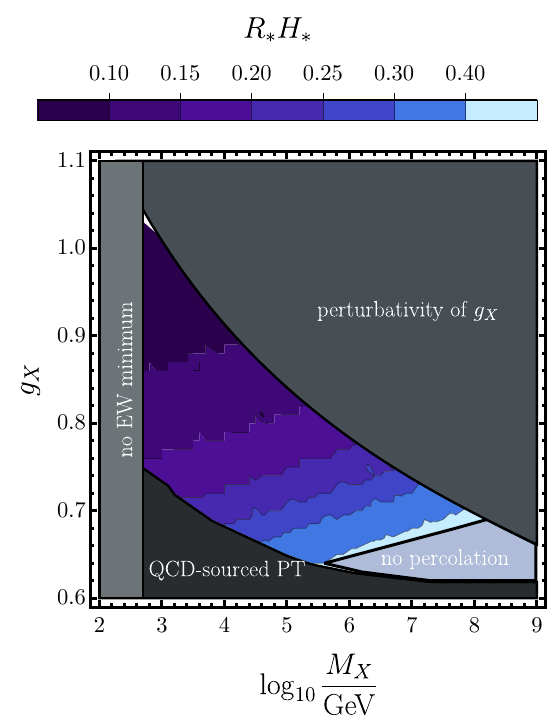}\\
\includegraphics[width=.4\textwidth]{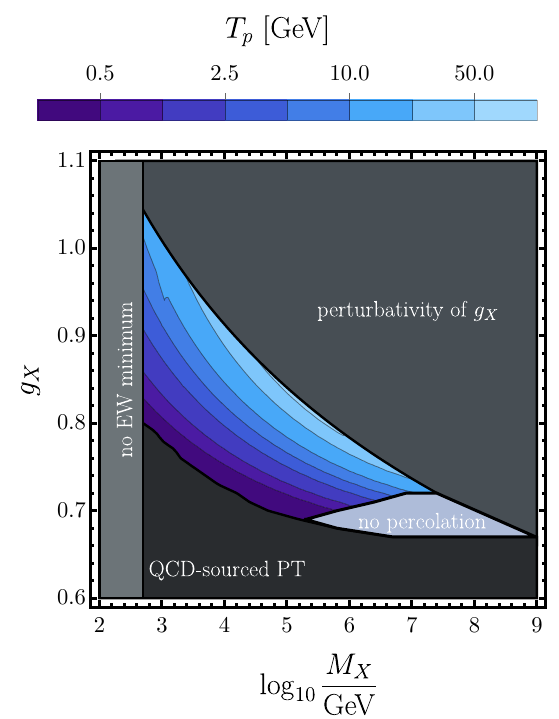}\hspace{2pt}
\includegraphics[width=.4\textwidth]{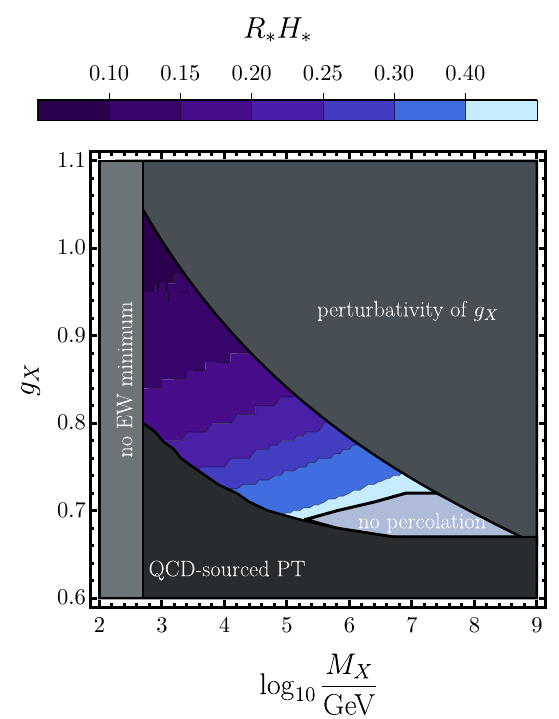}
\caption{Upper panel: Percolation temperature (left) and the transition length scale (right) obtained from the NLO potential with $\mu_{4} = \pi T$ and $\mu_3 =m_{X,3}^2(v_3)$. Lower panel: Percolation temperature (left) and the transition length scale (right) obtained from the LO potential with $\kappa$ in eq.~\eqref{eq:mu-choice} set to $\kappa=\pi$ to match the choice of $\mu_4$ for the NLO potential. The grey regions are excluded for reasons explained in the main text.}
\label{fig:Tp-R}
\end{figure}

Below we present the results of the scan of the parameter space obtained using the theory given by eqs.~\eqref{eq:Veff-EFT-LO},  \eqref{eq:VeftNLO}, \eqref{eq:Z3}. In figure~\ref{fig:Tp-R} (upper panel) we present the values of the percolation temperature and the average radius of a bubble at the moment of collision (length scale of the transition) obtained using the EFT NLO potential with $\mu_{4} = \pi T$. We exploit the invariance of the 
potential with respect to the 3D RG scale and choose a field-dependent value for it as $\mu_3=m_{X,3}^2(v_3)$ (see eq.~\eqref{eq:mass-gauge}). The excluded regions are marked in shades 
of grey. The leftmost part, corresponding to small $M_X$, does not reproduce the electroweak vacuum correctly. The upper right part of the parameter space is excluded as there the $g_X$ coupling 
becomes nonperturbative at some scale between $M_Z$ and the QCD scale $\mu\approx 0.1$\ GeV. In the lower part, the phase transition is triggered by the QCD condensate~\cite{Iso:2017uuu, vonHarling:2017, Marzola:2017, Hambye:2018, Baldes:2018,Ellis:2020, Sagunski:2023ynd, Wong:2023qon}.
We assume this happens for $T_p<0.1$\ GeV and this region is beyond the scope of the present 
work. Finally, in the light-grey triangle-shaped region we cannot assure the completion of the phase 
transition as the percolation criterion is not fulfilled. It was shown that supercooled phase transitions 
can produce primordial black holes (PBHs)  \cite{Gouttenoire:2023naa, Lewicki:2023ioy}, and for 
sufficiently slow transitions i.e.\ $\beta /H_* \lesssim 6-8 $ it would cause PBHs overabundance. We 
indeed find such values, but only in the right corner of the non-percolation region, for large masses 
$M_X$.
The former two constraints do not depend on the temperature, therefore they will be identical in all the plots. On the other hand, the latter two (regions where QCD effects become relevant and the percolation criterion is not satisfied) depend on our predictions for the phase transition and can change depending on the approach. In the lower panel of figure~\ref{fig:Tp-R} the same quantities computed from the LO potential of eq.~\eqref{eq:Veff-EFT-LO} (with matching conditions truncated as in eqs.~\eqref{eq:lambda3-LO}--\eqref{eq:m3sq-lo} to match the usual daisy-resummed approach) potential are presented. 

The values of the percolation temperature obtained at NLO are between 0.1 GeV and about 380 GeV. As a general trend, it can be noticed that the percolation temperature goes up, as compared to the LO prediction. This extends the available parameter space to lower values of the gauge coupling $g_X$. Moreover, the region of non-percolation is pushed to higher values of the $X$ mass. This opens up a small region of the parameter space where percolation is possible, but reheating is inefficient. We do not study this effect in detail as it is beyond the scope of the present paper.
\begin{figure}[!ht] 
\centering
\includegraphics[width=.45\textwidth]{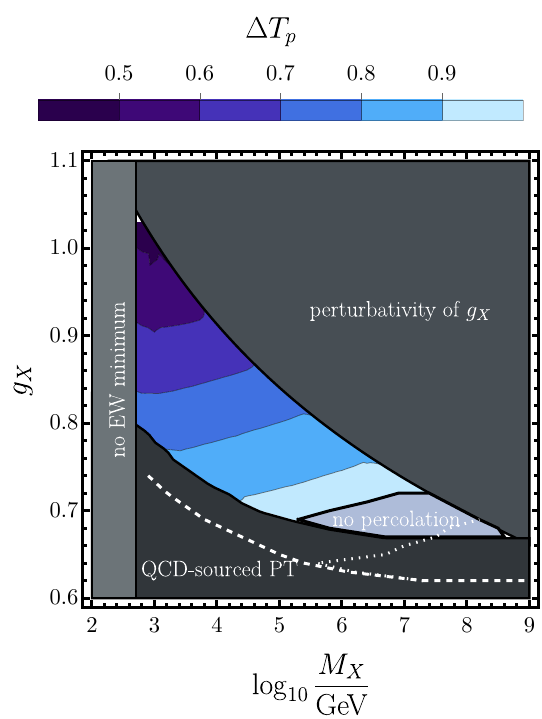}\hspace{2pt}
\includegraphics[width=.45\textwidth]{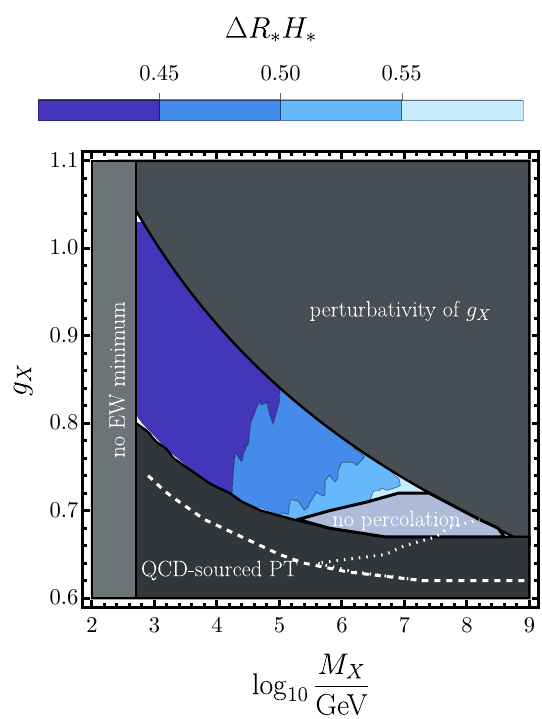}
\caption{Absolute value of the differences in the predictions for $T_p$ (left) and $R_*H_*$ (right) between the NLO and LO potentials, normalised by the NLO quantities. The excluded regions are the same as in figure~\ref{fig:Tp-R} (lower panel). The white dashed and dotted lines indicate the excluded regions obtained at NLO (as presented in the upper panel of figure~\ref{fig:Tp-R}).
}
\label{fig:Tp-R-diff}
\end{figure}

To better evaluate the differences between the LO and NLO predictions we present the relative differences between them (normalised to the NLO ones) in figure~\ref{fig:Tp-R-diff}. We can see that the change in the percolation temperature between the two approaches is significant, ranging from $\mathcal O(50\%)$ in the low-mass, large-coupling region, up to $\mathcal O(100 \%)$ in the small-couplings, large-mass corner. This seems somewhat counter-intuitive, as one expects the largest corrections between the two methods for \emph{large} couplings, but note that this concerns the coupling at the thermal scale. It should be stressed here that the coupling and mass displayed in figure~\ref{fig:Tp-R-diff} are defined at the scale $\mu = M_X$. They need to be RG-evolved to the thermal scale. In the large-mass corner, the coupling becomes significantly larger at the thermal scale, which explains why the difference between the LO and NLO approaches is largest in this part of parameter space. 
Let us point out again that the value of $T_p$ does not directly affect the GW spectrum, so the large corrections we find in the NLO description are not reflected in a strong modification of the GW signal. However, it signals that the differences between the descriptions at different orders in perturbation theory are non-negligible.

The right panel of figure~\ref{fig:Tp-R-diff} displays the difference in the typical length scale 
of the transition, $R_* H_*$. Here we see that the differences are smaller than for $T_p$, but again the largest
differences are observed in the large-mass corner, reaching $\mathcal O(55 \%)$.
Even though the relative difference between LO and NLO for $R_*H_*$ is smaller than for $T_p$, these differences \emph{do} modify the GW prediction. It can be seen in refs.~\cite{Kierkla:2022odc, Lewicki:2022pdb} that both the
GW amplitude and the peak frequency depend on  $R_*H_*$ and we, therefore, expect the predicted spectra to be shifted compared to the earlier results. Given the values of $R_*H_*$ we expect the GW signals to still be well visible in LISA. This will be verified by computing the SNR.

One can also note the change in the overall allowed region indicated in figure~\ref{fig:Tp-R-diff} by the white dashed and dotted lines. The region of non-percolation is significantly shifted and also the region where the phase transition is expected to be sourced by QCD effects is pushed to lower values of $g_X$ at NLO. Therefore, the predictions for the GW signal in this region could be significantly altered by using the LO or NLO approach.

\begin{figure}[t!] 
\centering
\includegraphics[width=.45\textwidth]{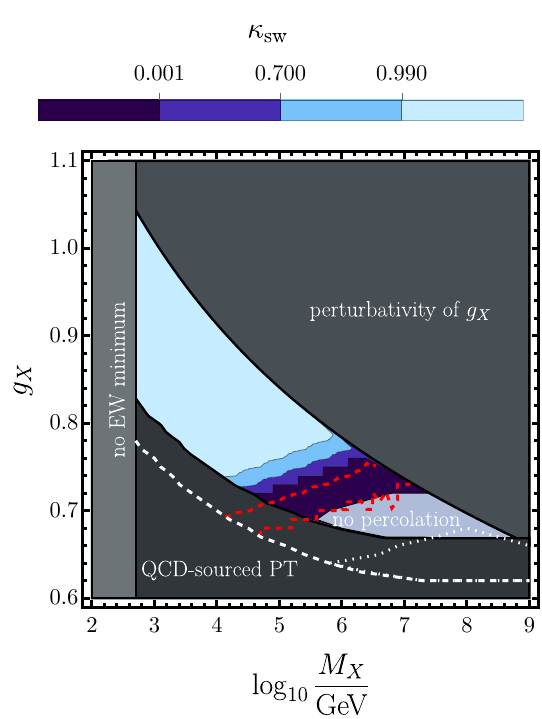}
\caption{
	The efficiency factor for sound waves $\kappa_{\textrm{sw}}$ computed from the LO potential. White lines correspond to excluded regions of parameter space obtained in the NLO setup. The red lines correspond to contours of $\kappa_{\textrm{sw}}=0.99$ (upper) and  $\kappa_{\textrm{sw}}=0.01$ (lower), obtained with the NLO potential.
}
\label{fig:kappa}
\end{figure}

To formulate the predictions for the GW spectra, we should ask the question about the source of the gravitational waves -- for which region of the parameter space are they produced via collisions of bubbles and where by sound waves?%
\footnote{The recent ref.~\cite{Lewicki:2022pdb} pointed out, based on numerical simulations, that for strongly supercooled transitions the spectra sourced by sound waves in the plasma look the same as the ones sourced by bubble collisions. Taking this into account, in practice the source of GW does not matter for the signal. It is anyway interesting to see, where in the parameter space supercooling is strong enough to allow for a runaway scenario.}
Our predictions for the efficiency factor for producing GWs via sound waves, $\kappa_{\mathrm{sw}}$ based on the LO and NLO potential are shown in figure~\ref{fig:kappa}.
It is clear that, depending on the region of the parameter space, GWs can be sourced by either of the mechanisms, and the difference in the predicted source between the two methods is limited to a rather narrow range of the parameter space. This is interesting, since in ref.~\cite{Kierkla:2022odc} for scans performed at a fixed renormalisation scale no region was found where the dominant source would correspond to bubble collisions (as opposed to predictions based on the RG-improved potential). In our current approach, the scale is fixed in the HT regime, where the tunnelling takes place. However, it is fixed to $\kappa T$ so effectively it is different for every point in the parameter space and is proportional to the percolation temperature. This suggests, that allowing the scale to change is crucial for seeing bubble collisions.%
\footnote{
Compare with the approach of ref.~\cite{Salvio:2023qgb}, where the scale was fixed and bubble collisions were assumed to be the main source of GW.
}

\begin{figure}[!ht] 
\centering
\begin{subfigure}[b]{0.45\textwidth}
	\includegraphics[width=\textwidth]{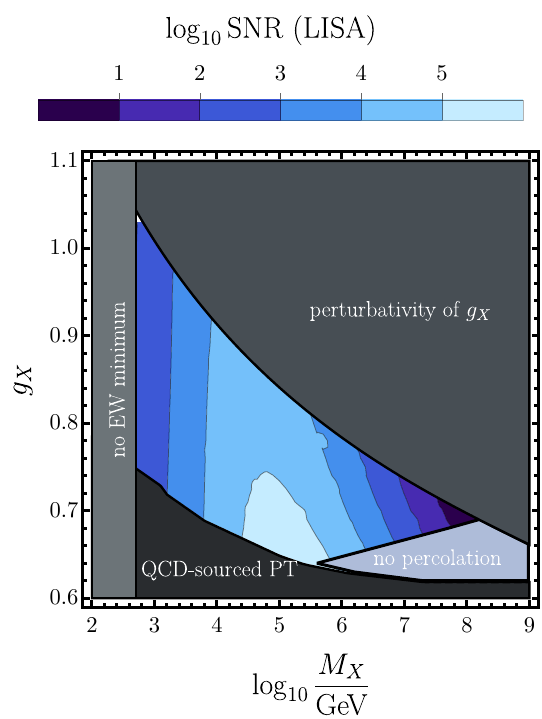}
\end{subfigure}
\begin{subfigure}[b]{0.45\textwidth}
	\includegraphics[width=\textwidth]{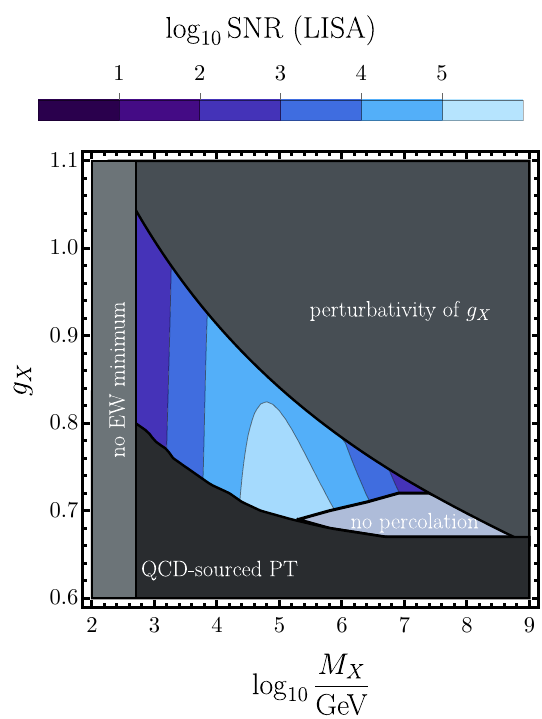}
\end{subfigure}
\caption{The values of SNR predicted at NLO (left panel), and at  LO (right panel). }
\label{fig:SNR}
\end{figure}

Having checked the predictions for the parameters of the transition, let us check what are the implications for the signal-to-noise (SNR) ratio predicted for LISA. The results of the LO and NLO approach are presented in figure~\ref{fig:SNR}. As expected, the values of SNR are very high throughout the entire allowed parameter space, implying that a first-order PT in the SU(2)cSM model should be well visible at LISA.%
\footnote{
	The values of the SNR were obtained using the approach described e.g. in \cite{Caprini:2015}. This approach is based on the assumption that the so-called ``self-noise'' of the GW signal is negligible. For the SU(2)cSM model, the GW signal is generically strong, therefore the obtained SNR values may be overestimated. We thank Kai Schmitz for pointing this out. Let us stress that the goal of this work is to emphasise the implications of higher order thermal corrections and the detailed treatment of the SNR is beyond the scope of this work. 
}
At NLO we observe a slightly lower SNR, around 10, at the edge of the parameter space 
corresponding to large $M_X$. The reason is that the peak frequency of the spectrum is higher for 
higher reheating temperatures. The latter grows with $M_X$ so for the largest values of $M_X$ 
allowed at NLO (excluded at LO by the percolation criterion), the signal moves out of the sensitivity 
range of LISA.

The very high values of SNR imply that, if a signal from a phase transition in SU(2)cSM is observed, we will be able to reconstruct the values of $T_r$ and $R_*H_*$ with very good precision, for very strong signals even better than 1\%~\cite{Gonstal:2023}.
Because of this prospect, the differences in predictions between the LO and NLO approaches, see figure~\ref{fig:Tp-R-diff}, will be much greater than the experimental uncertainties and this highlights the importance of
efforts to include higher-order thermal corrections.

\subsection{Implications for dark matter abundance}
The SU(2)$_X$ model contains DM candidates -- the new gauge bosons are stable due to a residual 
SO(3) symmetry and can in principle attain the correct relic density in different parts of the 
parameter space via the freeze-out 
mechanism~\cite{Khoze:2014,Pelaggi:2014wba,Karam:2015,Plascencia:2016,Hambye:2018,Baldes:2018,Marfatia:2020,Kierkla:2022odc,
 Frandsen:2022klh} or the super-cool DM 
mechanism~\cite{Hambye:2018,Baldes:2018,Marfatia:2020,Kierkla:2022odc}. The DM 
phenomenology is not the focus of the present paper but we will comment on how the NLO 
modifications of the allowed parameter space affect the existing results. In the lower-mass regime, 
the DM relic abundance is produced via the thermal freeze-out mechanism and a rather large gauge 
coupling is required. Reference~\cite{Kierkla:2022odc} showed that the correct relic abundance, in 
agreement with the current direct-detection limits, can be attained for $1.2\textrm{\ TeV}\lesssim 
M_X\lesssim 1.8\textrm{\ TeV}$ and $0.82\lesssim g_X\lesssim 0.96$ (similar results were found in 
ref.~\cite{Frandsen:2022klh}). The NLO results extend the allowed parameter space, where a 
first-order phase transition happens independently of QCD effects, to lower values of $g_X$ so we 
do not expect additional constraints on the region with the correct thermal DM abundance presented 
in 
refs.~\cite{Khoze:2014,Pelaggi:2014wba,Karam:2015,Plascencia:2016,Hambye:2018,Baldes:2018,Marfatia:2020,Kierkla:2022odc,
 Frandsen:2022klh}. 

 On the other hand, the mechanism of super-cool DM requires inefficient reheating, such that the temperature of the Universe after the phase transition is below the decoupling temperature of the $X$ particles. This is realised for larger $M_X$, approximately above 3000\ TeV, and lower $g_X$ around 0.7 (see e.g.\  figure~8 of ref.~\cite{Kierkla:2022odc}). This region was excluded by the LO analysis of ref.~\cite{Kierkla:2022odc}, however, the NLO results re-open a small part of this regime, as discussed in the previous subsection. Therefore, a small region with the correct relic density obtained via the super-cool DM mechanism (supplemented by a subthermal population produced via scattering) may be possible. However, to give a definite answer requires a dedicated computation of the reheating temperature and the resulting DM abundance.

\subsection{Evaluation of the uncertainties}
In section~\ref{sec:validityEFT} and~\ref{sec:scale-shifters} we have discussed the challenge of 
constructing an accurate EFT in a theory with scale-shifting fields, and the inaccuracy associated with the contribution from the bubble tail. 
Moreover, there is an uncertainty associated with the omission of higher-order corrections, which we can study by varying the RG scale.
In this section, we quantify the uncertainties associated with our computation of the thermal parameters.

\subsubsection{Dependence on the renormalisation scale}\label{sec:RGscaleDep}

As was explained earlier, the common approach of using the one-loop effective potential with daisy resummation suffers from an uncalled RG-scale dependence that can be cured by the inclusion of certain two-loop level diagrams. This is achieved in our NLO effective potential (with the matching also performed at two-loop level). There, the RG-scale dependence cancels up to terms of order higher than the order to which we compute. Moreover, the potential (and the full action) are independent of the 3D scale, up to higher-order corrections. As was shown in the literature~\cite{Croon:2020cgk,Gould:2021},
omission of significant perturbative corrections
(revealed by the scale dependence) is the main source of uncertainty in predicting the GW signals. We are now in a position to check the RG-scale sensitivity of the NLO predictions and to contrast it with the LO result. 
Note, that with the RG-improvement procedure implemented in this work, when we say that we change the 4d scale, in fact we mean that we change the thermal cutoff in the running in eq.~\eqref{eq:mu-choice}. So changing the scale from $\pi T$ to $2 \pi T$ means changing $\kappa$ in eq.~\eqref{eq:mu-choice} from $\kappa=\pi$ to $\kappa=2\pi$.
\begin{figure}[!t] 
\centering
\includegraphics[width=.45\textwidth]{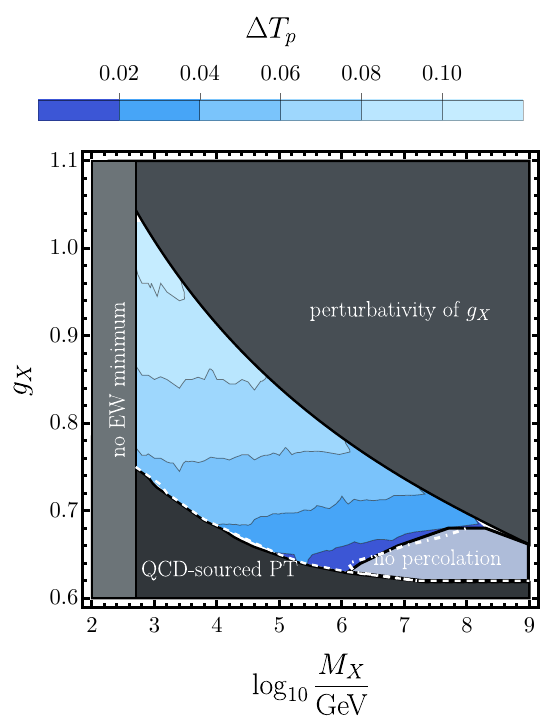}\hspace{2pt}
\includegraphics[width=.45\textwidth]{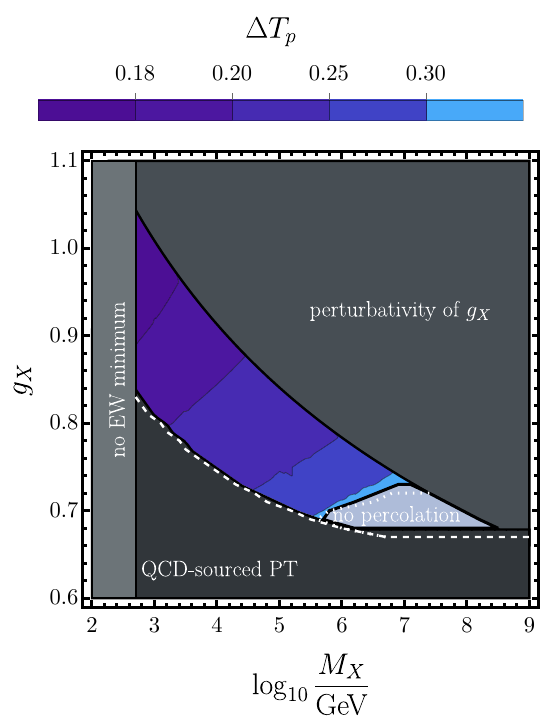}
\caption{Left panel: Absolute value of the differences in $T_p$ obtained from the NLO action with $\mu_4=\pi T$ and $\mu_4=2\pi T$, normalised by the result with $\mu_4=\pi T$. Right panel: Absolute value of the differences in $T_p$ obtained from the LO action with $\kappa=\pi$ and $\kappa=2\pi$, normalised by the result with $\kappa=\pi$. }
\label{fig:4d-scale-dep}
\end{figure}

Figure~\ref{fig:4d-scale-dep} (left panel) presents the relative difference in $T_p$ obtained from the NLO action at two different scales, $\mu_4=\pi T$ and $\mu_4=2\pi T$. We observe a mild dependence on the 4D scale, the result for $T_p$ changing, between the two RG-scales, by at most 10\%. The changes in $R_*H_*$ are much smaller, and they never exceed $2\%$, therefore we do not show the plot illustrating this difference. We have seen that the predicted SNR for LISA for this model is large, implying that thermal parameters can be reconstructed with very good precision. It therefore needs to be determined if the $2\%$ uncertainty in $R_*H_*$ leads to an observable difference.

For comparison, figure~\ref{fig:4d-scale-dep} (right panel) presents the dependence on the scale of the LO results. We can see that the change in results for $T_p$ is much larger -- the relative difference is approximately between 15\% and 30\%. This confirms our earlier claims that the inclusion of the NLO corrections cancels the residual scale dependence present at LO. The RG-scale dependence of the bubble radius is again milder and is of the order of 5\% in the whole parameter space, which is again larger than the uncertainty in the NLO result.

In both approaches, LO and NLO, the overall allowed region is only slightly modified by changing the 4D scale as indicated in figure~\ref{fig:4d-scale-dep} by the white dashed and dotted lines.

\subsubsection{Importance of the $Z_3$-factor and the bubble tail}
\label{sec:tail}

\begin{figure}[!ht] 
\centering
\includegraphics[height=.31\textwidth]{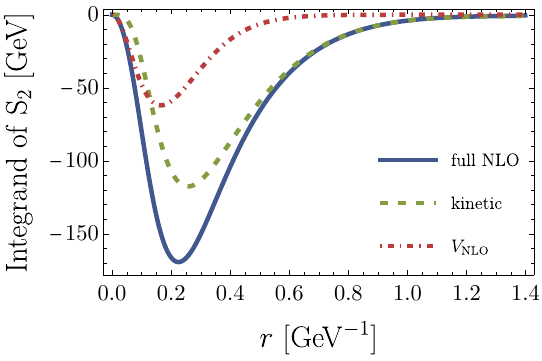}
\caption{Different contributions to the integrand of the NLO action $S_2$ of eq.~\eqref{eq:SbounceNLO} for $g_X = 0.8$, $M_X = 10$\,TeV, and $T_n= 14.18$\,GeV. The blue solid line  represents the full
NLO contribution, the red dot-dashed line the contribution from the NLO contribution to the potential and the dashed green line the contribution from the kinetic term with $Z_3$. }
\label{fig:NLOBounceAction}
\end{figure}

Looking at the rather good agreement between the different potentials in figure~\ref{fig:comparison-potentials}, the sizeable differences in $T_p$ and $R_*H_*$ observed in section~\ref{sec:GWs} might come as a surprise.
It turns out, that the largest cause of the difference between the LO and NLO descriptions is the $Z_3$-factor multiplying the kinetic term. As observed from its explicit form in eq.~\eqref{eq:Z3}, $Z_3$ diverges as $v_3 \rightarrow 0$. This corresponds to the regime where the
derivative expansion of the action is no longer valid, and
which our description hence cannot capture, see ref.~\cite{Ekstedt:2021kyx}.
Since along the bounce trajectory, as $v_3(r) \rightarrow 0$ also $\partial_i v_3(r) \rightarrow 0$, and $S_2$ thus remains finite. However, by comparing different contributions to $S_2$ shown in  figure~\ref{fig:NLOBounceAction},
we can see that the contribution of $Z_3$ is still dominant in $S_2$. 

To quantify the effect of $Z_3$ we have computed the thermal parameters in the mixed method with the NLO contribution to the potential, but without the $Z_3$ for a representative set of parameter choices. In all cases, we observe that the NLO approach gives a correction to $T_p$ only of the order of $\sim 25\%$ compared to the LO result, which is much smaller than the corrections observed in figure~\ref{fig:Tp-R-diff} and this confirms explicitly that the main correction comes from $Z_3$. 

A related question, which was already discussed in section~\ref{subsec:EFTs} and~\ref{sec:nucleation-in-EFT}, is the fraction of the action coming from the region with $r > r_t$. A large contribution from this region signals several problems. First, the solution gets a large contribution from the $Z_3$-factor in a region where its expression is not valid. Second, the EFT breaks down, due to the scale shifting nature of the fields. Last, large contributions coming from the kinetic term suggest that the derivative expansion, which allowed us to compute the bounce in a momentum-independent potential background, breaks down. In section~\ref{sec:nucleation-in-EFT} an approach was suggested to estimate the contribution from the region $r> r_t$. For the 3D RG-scale given by  $\mu_3 = m_{X,3}$, we find for a representative set of benchmark points that the fraction of the action given by $r> r_t$ is of the order $30-40\%$.
For the benchmark point considered in figure~\ref{fig:NLOBounceAction}, $r_t =0.29 \, {\rm GeV}^{-1}$ and the tail's contribution to the action is 31\%.
This is a significant fraction, and it should motivate us to investigate the validity of our expansion further. However, this estimate should not be taken too literally, as it depends on the choice of $\mu_3$. The fraction can be made smaller, at the cost of increasing $S_2$ with respect to $S_0$.

The way forward is to include the contributions of the gauge and scalar modes in the functional determinant of the nucleation rate, following the approach of refs.~\cite{Ekstedt:2021kyx,Ekstedt:2023sqc}. This will allow us to assess better the validity of the derivative expansion, and we leave it for future work.

\section{Summary and outlook}
\label{sec:summary}

This work is devoted to the accurate theoretical description of supercooled phase transitions, with the SU(2)cSM model~\cite{Hambye:2013, Carone:2013} as a concrete example. The main motivation for this study is the great prospect for observability of a GW signal from supercooled first-order phase transitions~\cite{Hambye:2013, Jaeckel:2016, Hashino:2016, Jinno:2016, Marzola:2017, Hashino:2018, Baldes:2018, Prokopec:2018,Marzo:2018, Mohamadnejad:2019vzg, Kang:2020jeg, Mohamadnejad:2021tke, Dasgupta:2022isg}, which would be so strong that spectral parameters could be reconstructed with a very good accuracy~\cite{Gonstal:2023}. This exciting possibility calls for an increased accuracy in the theoretical description, which is not accessible with the popular machinery of the daisy-resummed one-loop effective potential. A perfect tool for increasing the accuracy of phase-transition-related predictions is dimensionally reduced effective field theory~\cite{Ginsparg:1980ef,Appelquist:1981vg,Kajantie:1995dw,Braaten:1995cm} which allows us to perform resummations systematically. Therefore, we pursued the task of reconciling DR, which is based on a high-temperature expansion, with the description of supercooled phase transitions. This work resulted in several new findings which are summarised below. 

In section~\ref{sec:HT-LT-regimes} we demonstrated that the relevant quantities, describing the phase transition and setting the GW signal, can be divided into two groups: large-field-related ($\Delta V$, $T_r$) and small-field-related ($T_n$, $T_p$, $R_*H_*$). The former ones correspond to the low-temperature limit of the effective potential and no resummations are needed to compute them and we can follow the approach described e.g.\ in ref.~\cite{Kierkla:2022odc}. The latter ones are related to the high-temperature regime. To compute them accurately we need a high-temperature effective field theory which takes into account the hierarchies between different energy scales in the presence of high temperatures.

As we studied the relation between the HT and LT regimes, we also elucidated certain points in the computation of the thermal parameters of the phase transition in the 4D approach. First, by studying the interplay of the RG-improvement of the potential, and RG-scale cancellations between the HT limit of the thermal contribution to the effective potential and the zero-temperature part, we came to the conclusion that there is a preferred scale for the phase transition computations in the 4D theory which is the thermal scale, $\mu\sim T$, see section~\ref{sec:HT-LT-regimes}. This may sound like a trivial observation, but it is not commonly employed in the computations found in the literature. Moreover, we explained the role of the normalisation of the field in the phase-transition-related computations in section~\ref{sec:rescaling-field-4D}.

Furthermore, we have checked that the bounce solution, corresponding to the tunnelling trajectory of the field, is always within the HT regime. However, as we move along the bounce, the considered masses change substantially, which causes difficulties in treating mass hierarchies. The constructed EFT is expected to break down in the tail region, which is the region where the gauge field mass becomes smaller than the scalar field mass. In this region, contributions of the gauge field modes to the action should be accounted for without derivative expansion. These issues were discussed in sections~\ref{sec:validityEFT},~\ref{sec:scale-shifters} and~\ref{sec:tail}.

In sections~\ref{subsec:EFTs} and~\ref{sec:higher-orders} we have constructed an EFT in the HT regime, with matching at two-loop level and going to the NLO level in the couplings. This is the first time that these higher-order corrections are taken into account for a classically conformal model. It should be emphasised that at NLO, besides new contributions to the effective potential, there are also new effective operators that contribute to the kinetic part of the action, described by $Z_3$. This contribution can straightforwardly be included in the EFT framework, but is absent in the typical daisy-resummed approach. This is a serious shortcoming of the standard approach, as we find that the effect of $Z_3$ is significant. The significance of the $Z_3$-factor modifying the kinetic term as well the presence of the scale-shifting~\cite{Gould:2021ccf} fields suggests that the derivative expansion of the effective action might not be fully reliable. This should be studied by computing the functional determinant in the action prefactor, using the methodology developed in refs.~\cite{Ekstedt:2021kyx,Ekstedt:2022tqk,Ekstedt:2022ceo,Ekstedt:2023sqc}.

We formulated predictions based on the theory at NLO, by implementing a gauge-invariant and 3D 
RG-scale-invariant approach in section~\ref{sec:nucleation-in-EFT}, based on~\cite{Gould:2021ccf}. 
In such a setting, we performed a scan of the full parameter space of a BSM model, which has not 
yet been done in the context of GW production.%
\footnote{
Ref.~\cite{Friedrich:2022cak} is the first study that performs similar scan of a full BSM model parameter space in a gauge invariant formulation, yet still lacks 3D RG-scale invariance. 
}
We show in section~\ref{sec:GWs} that the differences in the percolation temperature $T_p$ between the LO and NLO approach become as large as $\mathcal O(1)$, with the largest corrections occurring in the large-mass, small-coupling corner. The differences in the predicted length scale of the transition $R_* H_*$ are more moderate, but also more relevant for the GW prediction, which depends strongly on $R_* H_*$.

We thoroughly study the scale dependence of the NLO predictions and compare them to the LO results in section~\ref{sec:RGscaleDep}. We find that the dependence on the RG-scale of the 4D theory, which is a measure for inaccuracy associated to missing higher-order corrections (see e.g.~\cite{Croon:2020cgk,Gould:2021}), becomes reduced in the NLO prediction compared to the LO prediction, indicating that higher-order corrections indeed are required to reduce this source of uncertainty. 

To sum up, we have demonstrated that higher-order corrections in the computation of thermal parameters in theories with supercooling \emph{can} and \emph{should be} included. We have found that the higher-order corrections have a significant effect on the GW signal, and that further studies into the contribution of the scale-shifting nature of the gauge fields is required.

\section*{Acknowledgements}

We thank 
Andreas Ekstedt,
Oliver Gould,
Joonas Hirvonen,
Johan L\"{o}fgren,
Lauri Niemi, 
Tomislav Prokopec
and Philipp Schicho
for illuminating discussions.
TT and JvdV thank University of Warsaw for hospitality during the initial stages of this project.
The work of B{\'S} and MK is supported by the National Science Centre, Poland, through the SONATA project number 2018/31/D/ST2/03302. 
MK is supported by the Polish National Agency for Academic Exchange under agreement PPN/PPO/2020/1/00013/U/00001 within the Polish Returns Programme.
JvdV is supported by the Dutch Research Council (NWO), under project number VI.Veni.212.133.
The work of TT has been supported in part by grants from the National Natural Science Foundation 
of China (grant nos. 11975150) from the Ministry of Science and Technology of China (grant no. 
WQ20183100522).

%

\appendix
\renewcommand{\thesection}{\Alph{section}}
\renewcommand{\thesubsection}{\Alph{section}.\arabic{subsection}}
\renewcommand{\theequation}{\Alph{section}.\arabic{equation}}

%
\section{Running couplings and field \label{app:betas}}

The $\beta$ functions for the scalar couplings $\lambda_{\field}$, $\lambda_{h\field}$, $\lambda_h$ and the new gauge coupling $g_X$ read~\cite{Khoze:2014, Carone:2013, Hambye:2013}
\begin{displaymath}
\begin{aligned}
\beta_{\lambda_h} &= \frac{1}{8\pi^2}\left[12\lambda_h^2+\lambda_{h\field}^2+\frac{\lambda_h}{2}\left(-9g_2^2-3g_1^{ 2}+12Y_t^2\right)+\frac{3g_2^4}{8}+\frac{3(g_2^2+g_1^{ 2})^2}{16}-3Y_t^4\right],\\
\beta_{\lambda_{h\field}} &= \frac{1}{8\pi^2}\left[6\lambda_h\lambda_{h\field}+2\lambda_{h\field}^2+6\lambda_{h\field}\lambda_{\field}+\frac{\lambda_{h\field}}{4}\left(-9g_2^2-3g_1^{ 2}+12Y_t^2-9g^2_{X}\right)\right], \\
\beta_{\lambda_{\field}} &= \frac{1}{8\pi^2}\left[\lambda_{h\field}^2+12\lambda_{\field}^2-\frac{9}{2}\lambda_{\field}g^2_{X}+\frac{9g^4_{X}}{16}\right],\\
\beta_{g_X} &= \frac{1}{16\pi^2}\left[-\frac{43}{6}g^3_{X} - \frac{1}{(4\pi)^2}\frac{259}{6}g^5_{X}\right],
\end{aligned}
\end{displaymath}
where $g_2$ is the SU(2) gauge coupling, $g_1$ the U(1) gauge coupling and $Y_t$ the top Yukawa coupling. The 4D renormalisation factor for the scalar field $\field$ is given by
$$
Z(t)=\exp\left(- \frac{1}{2}\int_0^t \dd x\ \gamma_{\field}(x)\right),
$$
where  $t=\log\frac{\mu}{\mu_0}$ and
$$
\gamma_{\field}(x)=-\frac{9 g_X^2(x)}{32 \pi ^2}.
$$
These $\beta$ functions are also obtained automatically when using {\tt DRalgo}~\cite{Ekstedt:2022bff} for dimensional reduction.

%

\section{Dimensional reduction}
\label{sec:DR-details}
Customarily in the dimensional reduction literature, the scale of the phase transition is referred to as the ultrasoft scale and the corresponding EFT is the one where the soft scale temporal gauge  field components are integrated out~\cite{Kajantie:1995dw}, along with other potential soft fields~\cite{Niemi:2018asa}.  As pointed out in ref.~\cite{Gould:2023ovu}, only in the near vicinity of a second-order phase transition, the scalar fields are expected to become ultrasoft. For first-order transitions, the fields driving the transition are expected to live either at the supersoft or the soft scales.
In this article, we have indeed organised perturbation theory by treating the nucleating field  as soft and constructed the potential and nucleation EFT in the broken phase by integrating out the gauge fields, both spatial and temporal components on equal footing. In section~\ref{sec:dralgo} we demonstrate how to obtain the effective theory using {\tt DRalgo}, and in section~\ref{sec:Matching} we list the relevant matching relations.

\subsection{Using \tt{DRalgo}}
\label{sec:dralgo}
For completeness, we describe here our implementation in {\tt DRalgo}. We refer to ref.~\cite{Ekstedt:2022bff} for details of the program. Our results were obtained using version 1.01 beta (16-05-2022) of {\tt DRalgo} and version 1.1.2 (06-05-2020) of {\tt Groupmath}~\cite{Fonseca:2020vke}.

We start from defining the model of section~\ref{sec:model} as follows:
\begin{lstlisting}[language=Mathematica,mathescape=true]
Group = {"SU3", "SU2", "SU2", "U1"};
RepAdjoint = {{1, 1}, {2}, {2}, 0}; (* gauge fields in adjoint representations *)
HiggsDoublet1 = {{{0, 0}, {1}, {0}, 1/2}, "C"}; (* Higgs doublet with hypercharge 1/2 *)
HiggsDoublet2 = {{{0, 0}, {0}, {1}, 0}, "C"}; (* dark sector doublet with zero hypercharge *)
RepScalar = {HiggsDoublet1, HiggsDoublet2};
CouplingName = {g3, g2, gX, g1};

(* SM fermion content per generation: *)
Rep1 = {{{1, 0}, {1}, {0}, 1/6}, "L"}; (* L-handed quark doublet *)
Rep2 = {{{1, 0}, {0}, {0}, 2/3}, "R"}; (* R-handed up-type quark *)
Rep3 = {{{1, 0}, {0}, {0}, -1/3}, "R"}; (* R-handed down-type quark *)
Rep4 = {{{0, 0}, {1}, {0}, -1/2}, "L"}; (* L-handed lepton doublet *)
Rep5 = {{{0, 0}, {0}, {0}, -1}, "R"}; (* R-handed lepton *)
RepFermion1Gen = {Rep1, Rep2, Rep3, Rep4, Rep5};

(* 3 duplicate fermion generations: *)
RepFermion3Gen = {RepFermion1Gen, RepFermion1Gen, RepFermion1Gen} // 
   Flatten[#, 1] &;

(* allocate required tensors: *)
{gvvv, gvff, gvss, \[Lambda]1, \[Lambda]3, \[Lambda]4, \[Mu]ij, \[Mu]IJ, \[Mu]IJC, Ysff, YsffC} = AllocateTensors[Group, RepAdjoint, CouplingName, RepFermion3Gen, RepScalar];


(* construct bilinear scalar invariants: *)
InputInv = {{1, 1}, {True, False}}; (*\[Phi]1\[Phi]1^+*)
MassTerm1 = CreateInvariant[Group, RepScalar, InputInv] // Simplify;
InputInv = {{2, 2}, {True, False}}; (*\[Phi]2\[Phi]2^+*)
MassTerm2 = CreateInvariant[Group, RepScalar, InputInv] // Simplify;

(* Here we implement auxilliary mass terms with squared masses m1sq and m2sq, that we set to zero at the very end of computation *)
VMass = (m1sq*MassTerm1 + m2sq*MassTerm2);
\[Mu]ij = GradMass[VMass[[1]]] // Simplify // SparseArray;

(* construct quartic terms from the invariants: *)
QuarticTerm1 = MassTerm1[[1]]^2; (*[(\[Phi]1\[Phi]1^+)]^2*)
QuarticTerm2 = MassTerm2[[1]]^2; (*[(\[Phi]2\[Phi]2^+)]^2*)
QuarticTerm3 = MassTerm1[[1]]*MassTerm2[[1]]; (* (\[Phi]1\[Phi]1^+)(\[Phi]2\[Phi]2^+)^2*)

(* collect quartic terms: *)
VQuartic = (\[Lambda]h*QuarticTerm1 + \[Lambda]\[Psi]*QuarticTerm2 + \[Lambda]h\[Psi]*QuarticTerm3);
\[Lambda]4 = GradQuartic[VQuartic];

(* construct Yukawa interaction: *)
InputInv = {{1, 1, 2}, {False, False, True}}; 
YukawaDoublet1 = 
	CreateInvariantYukawa[Group, RepScalar, RepFermion3Gen, InputInv] // Simplify;
    
(* only top quark has non-zero Yukawa coupling: *)
Ysff = -yt*GradYukawa[YukawaDoublet1[[1]]];
YsffC = SparseArray[
   Simplify[Conjugate[Ysff] // Normal, Assumptions -> {yt > 0}]];

(* create the model: *)
ImportModelDRalgo[Group, gvvv, gvff, 
gvss, \[Lambda]1, \[Lambda]3, \[Lambda]4, \[Mu]ij, \[Mu]IJ, \
\[Mu]IJC, Ysff, YsffC, Verbose -> False];
\end{lstlisting}
Note that due to the design choice of {\tt DRalgo}, we need to define the model \emph{with} masses in the 4D theory ({\tt m1sq} and {\tt m2sq}), and then set them to zero later.

We obtain the couplings and masses of the effective theory by running 
\begin{lstlisting}[language=Mathematica,mathescape=true]
PerformDRhard[]
PrintScalarMass["NLO"] (* prints out the result for 2-loop scalar thermal masses *)
PrintDebyeMass["NLO"] (* prints out the result for 2-loop Debye masses *)
PrintCouplings[] (* prints out the result for 1-loop thermal corrections of couplings *)
PrintTemporalScalarCouplings[] (* prints out the result for temporal gauge field couplings *)
\end{lstlisting}
The obtained masses and couplings are listed in section~\ref{sec:Matching}, up to two modifications:
\begin{itemize}
\item{We obtain $\kappa_3$ from ref.~\cite{Kajantie:1995dw} rather than {\tt DRalgo}. 
The reason is that the normalisation convention of $\kappa_3$ in {\tt DRalgo} does not align with literature following ref.~\cite{Kajantie:1995dw}, e.g.\ refs.~\cite{Brauner:2016fla, Andersen:2017ika} (see Q.10 of {\tt DRalgo}'s Q\&A).}
\item{We remove the terms in $m_3^2$ that multiply $\ln{\left( \mu_3/\mu_4 \right)}$ which correspond to running due to modes that we did \emph{not} include in the NLO effective potential, which gets contributions only from the soft modes.
}
\end{itemize}
At the moment this computation was performed, {\tt DRalgo} did not allow the construction  of our NLO potential as the temporal gauge field is integrated out before higher-order corrections to the effective potential can be computed.%
\footnote{
A recent update of {\tt DRalgo} does allow the user to construct the NLO potential while treating the temporal and spatial gauge modes in the same way.
}
For the dark sector effective potential we used a trick to account for the contributions of the temporal gauge field $X^a_0$, by adding an additional scalar triplet under the dark SU(2):

\begin{lstlisting}[language=Mathematica,mathescape=true]
TemporalGaugeTriplet1 = {{{0, 0}, {0}, {2}, 0}, "R"}; (* a real dark triplet *)
RepScalar = {HiggsDoublet1, HiggsDoublet2, TemporalGaugeTriplet1};
\end{lstlisting}
In order to keep this dark SU(2) triplet field within the EFT that {\tt DRalgo} uses to compute the effective potential, we introduced it in the parent theory. This is merely a hack: we do not account for its contributions in the matching relations, as it is nothing but an auxiliary field, that we use to circumvent the limitations of vanilla {\tt DRalgo}. We define a mass term and a coupling term for the auxiliary field via
\begin{lstlisting}[language=Mathematica,mathescape=true]
MassTerm3 = CreateInvariant[Group, RepScalar, InputInv] // Simplify;
QuarticTerm4 = MassTerm3[[1]]^2;
QuarticTerm5 = MassTerm2[[1]]*MassTerm3[[1]];
\end{lstlisting}
and the corresponding tree-level potential becomes
\begin{lstlisting}[language=Mathematica,mathescape=true]
VMass = (
   	+ m1sq*MassTerm1
    	+ m2sq*MassTerm2
    	+ 1/2 mx0sq*MassTerm3 (* triplet mass term *)
   	);
   	
VQuartic = (
   	+\[Lambda]h*QuarticTerm1
    	+ \[Lambda]\[Psi]*QuarticTerm2
    	+ \[Lambda]h\[Psi]*QuarticTerm3
    + b4/4*QuarticTerm4 (* triplet self interaction *)
    + a2/2 QuarticTerm5 (* portal between triplet and dark doublet *)
   	);
\end{lstlisting}
Apart from these modifications, the model implementation is identical to the one described above.

The computation of the effective potential now proceeds as follows. First, the EFT at the soft scale is constructed via
\begin{lstlisting}[language=Mathematica,mathescape=true]
PerformDRsoft[{}]
\end{lstlisting}
We then specify that only the $\varphi$-field obtains a vev and compute the effective potential
\begin{lstlisting}[language=Mathematica,mathescape=true]
(* first 4 elements correspond to the SM doublet, next 4 dark doublet, and last 3 auxilliary triplet *)
\[CurlyPhi]vev = {0, 0, 0, 0, 0, 0, 0, w, 0, 0, 0} // SparseArray
DefineVEVS[\[CurlyPhi]vev]
CalculatePotentialUS[]
\end{lstlisting}
The tree-level effective potential at the soft scale (without the higher-dimensional corrections) of eq.~(\ref{Veff:softTree}) is obtained by calling
\begin{lstlisting}[language=Mathematica,mathescape=true]
PrintEffectivePotential["LO"] (* tree-level effective potential *)
\end{lstlisting}
We obtain the leading order effective potential of the final EFT, eq.~(\ref{eq:Veff-EFT-LO}), by adding the result of 
\begin{lstlisting}[language=Mathematica,mathescape=true] 
PrintEffectivePotential["NLO"] (* 1-loop effective potential *)
\end{lstlisting}
where we set $a_2 \rightarrow h_3$ and we set the scalar contributions to zero, as they originate from the parameterically lighter modes. We also add a field-independent term by hand which ensures that the potential is normalised to zero at the vanishing field value. Lastly, we obtain the NLO contribution to the effective potential, eq.~(\ref{eq:VeftNLO}), from 
\begin{lstlisting}[language=Mathematica,mathescape=true]
PrintEffectivePotential["NNLO"] (* 2-loop effective potential *)
\end{lstlisting}
where we again remove the scalar field contributions by setting $\lambda_{h}$ and $\lambda_{h\varphi}$ to zero. The parameter $b_4$ should correspond to $\kappa_3$, but since we used the normalisation of ref.~\cite{Kajantie:1995dw}, we read off the pre-factor of this term from ref.~\cite{Gould:2023ovu}, which uses the same convention.

\subsection{Hard to soft scale matching relations}
\label{sec:Matching}
We perform next-to-leading (NLO) matching when integrating out the hard modes. This corresponds to one-loop matching for the couplings and two-loop matching for the masses. The scalar field couplings are given by
\begin{align}
\lambda_{3} &= T \bigg( \lambda_{\varphi} + \frac{1}{(4\pi)^2} \Big( \frac{3}{8}g^4_X + L_b ( -\frac{9}{16} g^4_X - 12 \lambda_{\varphi} + \frac{9}{2} g^2_X \lambda_{\varphi} -  \lambda^2_{h\varphi} ) \Big) \bigg), \\
g^2_{X,3} &= g^2_X T \Big(1 + \frac{1}{(4\pi)^2}g^2_X ( \frac{2}{3} + \frac{43}{6}L_b) \Big),\label{eq:lambda3} \\
h_3 &= \frac{1}{2} g^2_X  T  \bigg(1 + \frac{1}{(4\pi)^2} \Big( g^2_X (\frac{17}{2} + \frac{43}{6}L_b) + 12 \lambda_{\varphi} \Big) \bigg), \\
\kappa_3 &= \frac{1}{(4\pi)^2} \frac{17}{3} g^4_X T,
\end{align}
and the masses by
\begin{align}
m^3_{D,X} &= \frac{5}{6} g^2_X T^2 \bigg( 1 + \frac{1}{(4\pi)^2} T^2 \Big( g^2_X (\frac{69}{20} + \frac{43}{6}L_b) + \frac{2}{5}(3\lambda_\varphi  +  \lambda_{h\varphi}) \Big) \bigg),\label{eq:mDX} \\
m^2_3 &= T^2 \bigg\{ \frac{3}{16} g^2_X + \frac{1}{6}(3 \lambda_\varphi + \lambda_{h\varphi} ) \nonumber \\
& + \frac{1}{(4\pi)^2} \bigg( \frac{167}{96} g^4_{X} + \frac{1}{12}(3g^2_{2} + g^2_{1} + 3 L_f y^2_t) \lambda_{h\varphi} + \frac{3}{4} g^2_{X} \lambda_{\varphi} \nonumber \\
& + L_b \Big( -\frac{47}{32} g^4_{X} - \frac{9}{4} g^2_{X} \lambda_{\varphi} - \frac{1}{8} (9 g^2_{2} + 3g^2_{1} + 3 g^2_{X} + 6 y^2 + 8  \lambda_{h} + 8  \lambda_{\varphi} ) \lambda_{h\varphi} + \frac{1}{6} \lambda^2_{h\varphi}  \Big) \nonumber \\
& + \gamma \Big( \frac{81}{32} g^4_{X} + \frac{1}{2} (3g^2_{2} + g^2_{1}) \lambda_{h\varphi} - \lambda^2_{h\varphi} + \frac{9}{2} g^2_{X} \lambda_{\varphi} - 6 \lambda^2_{\varphi} \Big) \nonumber \\ 
& + \ln(A) \Big( -\frac{243}{2} g^4_{X} - 6(3g^2_{2} + g^2_{1})\lambda_{h\varphi} + 12 \lambda^2_{h\varphi} - 54 g^2_{X} \lambda_{2} + 72 \lambda^2_{\varphi} \Big)  \bigg) \bigg\} \nonumber \\
& + \frac{1}{(4\pi)^2} \ln\Big( \frac{\mu_3}{\mu_4} \Big) \bigg( -\frac{36}{16}g^4_{X,3}  + \frac{3}{2} \lambda^2_{VL8} - 6 g^2_{X,3} \lambda_{VL8}  \bigg),\label{eq:3DMass}
\end{align}
where
\begin{align}
\lambda_{h\varphi,3} & = \lambda_{h\varphi}T\bigg(1 + \frac{1}{(4\pi)^2} \Big( -3 y_t^2 L_f + L_b(\frac{3}{4}g_1^2 + \frac{9}{4} g_2^2 + \frac{9}{4}g_X^2 - 2(3 \lambda_h + \lambda_{h\varphi} + 3 \lambda_{\varphi}) ) \Big)\bigg), \\
g_{1,3} &= \sqrt{g_1^2 T - \frac{g_1^4T(L_b + 40 L_f)}{96 \pi^2}}, \\
g_{2,3} &= \sqrt{g_2^2 T + \frac{g_2^4 T(4 + 43 L_b - 24 L_f)}{96 \pi^2}}.
\end{align}
$L_b$ and $L_f$ are defined as
\begin{equation}\label{eq:LbLf}
L_b = 2 \gamma_E - 2\log{[4\pi]} + \log{\left[\frac{\mu_4^2}{T^2} \right]}, \qquad L_f = L_b + 4\log{[2]},
\end{equation}
with $\gamma_E$ the Euler-Mascheroni's constant and $A$ the Glaisher's constant.

The effective masses depend on the effective couplings between the scalar and the several temporal gauge modes, given by
\begin{align}
\lambda_{VL5} &= \frac{g_1^2 T \lambda_{h\varphi}}{8\pi^2}, \\
\lambda_{VL6} &= \frac{g_2^2 T \lambda_{h\varphi}}{8\pi^2}, \\
\lambda_{VL8} &= \frac{g_X^2 T (g_X^2(51 + 43 L_b) + 96 \pi^2 + 72 \lambda_\varphi)}{1928\pi^2},
\end{align}
where $\lambda_{VL5},\lambda_{VL6}$ denote couplings between $\varphi$ and the SM gauge fields, and $\lambda_{VL8}$ the coupling between $\varphi$ and the dark gauge field.

{\small

\bibliography{conformal-bib+GW.bib}

}
\end{document}